\documentclass{revtex4-1}  

\usepackage{geometry}
\usepackage{graphicx}

\usepackage{amssymb}
\usepackage{amsmath}
\usepackage[utf8x]{inputenc} 
\usepackage{bm}

\usepackage{xcolor}
\usepackage[colorlinks=true, linkcolor=blue, citecolor = blue]{hyperref}

\usepackage{lineno}
\modulolinenumbers[1]
\usepackage{siunitx}
\usepackage{textcomp}
\usepackage{palatino}

\begin{document}

\title{Shape transition and hydrodynamics of vesicles in tube flow}

\author{Paul G. Chen}
\email{gang.chen@univ-amu.fr}
\affiliation{Aix Marseille Univ, CNRS, Centrale Marseille, M2P2, Marseille, France}
\author{J. M. Lyu}
\affiliation{Aix Marseille Univ, CNRS, Centrale Marseille, M2P2, Marseille, France}
\author{M. Jaeger}
\affiliation{Aix Marseille Univ, CNRS, Centrale Marseille, M2P2, Marseille, France}
\author{M. Leonetti}
\email{leonettm@univ-grenoble-alpes.fr}
\affiliation{Univ. Grenoble Alpes, CNRS, Grenoble INP, LRP, Grenoble, France}

\date{\today}					% Activate to display a given date or no date

\begin{abstract}

The steady motion and deformation of a lipid-bilayer vesicle translating through a circular tube in low Reynolds number pressure-driven flow are investigated numerically using an axisymmetric boundary element method. This fluid-structure interaction problem is determined by three dimensionless parameters: reduced volume (a measure of the vesicle asphericity), geometric confinement (the ratio of the vesicle effective radius to the tube radius), and capillary number (the ratio of viscous to bending forces).  The physical constraints of a vesicle -- fixed surface area and enclosed volume when it is confined in a tube -- determine critical confinement beyond which it cannot pass through without rupturing its membrane. The simulated results are presented in a wide range of reduced volumes [0.6, 0.98] for different degrees of confinement; the reduced volume of 0.6 mimics red blood cells.  We draw a phase diagram of vesicle shapes and propose a shape transition line separating the parachute-like shape region from the bullet-like one in the reduced volume versus confinement phase space. We show that the shape transition marks a change in the behavior of vesicle mobility, especially for highly deflated vesicles.  Most importantly, high-resolution simulations make it possible for us to examine the hydrodynamic interaction between the wall boundary and the vesicle surface at conditions of very high confinement, thus providing the limiting behavior of several quantities of interest, such as the thickness of lubrication film, vesicle mobility and its length, and the extra pressure drop due to the presence of the vesicle.  This extra pressure drop holds implications for the rheology of dilute vesicle suspensions. Furthermore, we present various correlations and discuss a number of practical applications.  The results of this work may serve as a benchmark for future studies and help devise tube-flow experiments.

\end{abstract}

\maketitle

\section{Introduction}

The transport of deformable particles (such as drops, vesicles, capsules, red blood cells, etc.) in microchannels has recently received much attention because of its key roles in applications ranging from microcirculation of red blood cells (RBCs)~\cite{Pries_1992,SECOMB2013470,Tomaiuolo2009,Quint2017} and biomimetic carriers~\cite{Europhysics_Vitkova_2004,Risso_JFM2006} to manipulating droplets in microfluidic chips \cite{Cantat2013,Huerre2015}. It is therefore of great interest to understand how those fluid-particles deform in response to external forces in narrow tubes and how these forces affect the hydrodynamic mobility of the deformable particles.  In addition to the particle's mobility~\cite{Europhysics_Vitkova_2004}, which is defined as the ratio of the particle velocity to the suspending mean fluid velocity, the other important quantity of interest in the context of blood rheology is the extra pressure drop due to the presence of the particle in the capillary tube~\cite{Pries_1992,SECOMB2013470}. Irrespective of the type of deformable particles, their shapes are unknown {\it a priori} due to a highly nonlinear fluid-structure coupling, and their dynamics are dictated by interfacial mechanics, boundary conditions on the walls, mechanical equilibrium at the interface, and other specific constraints (e.g., a fixed internal volume in general and a fixed surface area in particular for vesicles). 

A soft object transported in confined flows is essentially a fluid-structure interaction problem. A variety of  numerical models have been developed to predict the motion of a deformable entity. They  can be broadly classified into three categories, namely, mesh-based methods~\cite{Bagchi_2012,Mendez_2014,ZHANG2019Lattice}, particle-based methods~\cite{Noguchi2005Shape,FedosovCaswell2010b,Lanotte2016red}, and boundary-element method (BEM). We refer to recent review articles for a comprehensive overview of the literature (e.g., Ref.~\cite{Barthes_2016} for capsules, Ref.~\cite{Abreu_2014} for vesicles, and Ref.~\cite{Freund_2014} for RBCs).  The BEM is arguably the most popular simulation framework for highly accurate prediction of the dynamics of deformable particles in inertialess Newtonian flows. A major advantage of the BEM as compared with the domain-based numerical methods is that it reduces the spatial dimensions of the computational domain by one, so the flow equations are solved only for the unknown stress and velocity fields at the domain boundaries and at moving interfaces.  BEM's theory and formulation are well described in the book by Pozrikidis~\cite{Pozrikidis_1992}, and its efficiency has been demonstrated in the simulation of drops~\cite{Lac_JFM2009,Nagel2015Boundary}, capsules~\cite{Hu_JFM_2012}, vesicles~\cite{Boedec_JCP_2011,Farutin_JCP_2014,Boedec_JCP_2017,barakat_shaqfeh_2018b,barakat_shaqfeh_2019}, and blood cells~\cite{Pozrikidis_2005b,Zhao_JCP_2010,Quint2017}.

In this paper, we focus on the dynamical behavior of a confined axisymmetrical vesicle moving through a circular tube in pressure-driven flow. Membranes of vesicles ($\sim4$ nm thick) are made of a lipid bilayer that behaves as a two-dimensional fluid contrary to polymer capsules. Under stress, mechanical responses of membranes are resistance to an out-of-plane bending but not to an in-plane flow. In most experimental configurations, the lipid bilayer membrane is incompressible, a physical constraint ensured theoretically via a local Lagrange multiplier which has the same dimensions as surface tension (i.e., N/m or J/m$^2$).  In addition, vesicles are only deformable if they are initially deflated, a characteristic measured by reduced volume or excess area,  which explains why vesicles are singular particles~\cite{Seifert_PRA1991}. Vesicles are often employed as a model system to mimic the mechanical properties of red blood cells~\cite{Lipowsky_1991,Seifert_PRA1991,Vlahovska_2013}. 

Recent semi-analytical progress has been achieved in the limit of highly confined axisymmetrical vesicles in tube flow \cite{barakat_shaqfeh_2018a}.  It means that dissipation is mainly governed by the lubrication in the thin film between the vesicle and the tube wall.  However, the lubrication equation derived from the Reynolds equation and mechanical equilibrium (Eq. (5.27) in \cite{barakat_shaqfeh_2018a}) needs careful numerical integration as done by the authors. Asymptotic solutions are only possible for vanishing bending resistance -- which corresponds to an infinite bending capillary number, a measure of the relative importance of hydrodynamic stress compared to bending resistance. The most significant result of Ref.~\cite{barakat_shaqfeh_2018a} is a prediction of vesicle's mobility and extra pressure drop in situations where the gap size between the vesicle surface and the tube wall is small.
Calculation of vesicle shape and membrane stress in a confined configuration has begun with Trozzo et al.~\cite{Trozzo_JCP_2015} who developed an axisymmetric BEM for vesicles moving through a circular tube. But several hydrodynamical quantities of interest, such as vesicle's velocity and extra pressure drop, were not provided. Three-dimensional BEM simulations of confined vesicles have been performed recently in a range of reduced volumes $0.7$--$0.99$~\cite{barakat_shaqfeh_2018b}. The authors also presented an axisymmetric lubrication theory based on the parallel-flow approximation \cite{SecombSkakak1986}. This approximation is appropriate if the gap size is small compared to the vesicle length. A comparison of vesicles shapes computed via 3D BEM simulations and axisymmetric lubrication theory shows the relevance of the lubrication theory provided that the gap size and the reduced volume are not too small (Fig. 9 of \cite{barakat_shaqfeh_2018b}). Indeed, those highly confined regimes are difficult to reach numerically as an extremely refined meshing is necessary to capture the hydrodynamic interaction in the lubrication layer; thus, axisymmetric BEM simulations remain an interesting and promising approach. Also, it is expected that the vesicle preserves a steady, axisymmetric configuration when it is highly confined. 

Here, we study numerically the motion and deformation of an axisymmetric vesicle moving inside a tube in a wider range of the reduced volumes, i.e., $0.6$--$0.98$; the value of 0.6 corresponds to the reduced volume of red blood cells. We also compute explicitly the thickness of the lubrication film, which has not been considered in previous works. In Sec.~\ref{pb}, we present problem formulation and dimensionless parameters involved. In Sec.~\ref{Num}, we outline the numerical method used in this study. In Sec.~\ref{Res}, we present and discuss our main results. Firstly, we determine the vesicle shapes with such reduced volumes at varying degrees of confinement. Secondly, we focus on the transition between parachute shapes (concave rear like a droplet) and bullet shapes (convex rear) as first studied in free space for quasi-spherical vesicles under quadratic flow (i.e., Poiseuille flow but without walls) \cite{Farutin_PRE_2011}. Comparisons with experiments on single vesicles \cite{Europhysics_Vitkova_2004} and on the phase diagram \cite{Coupier_PhysRevLett108_2012} show very good agreement. Thirdly, we present numerical results of vesicle's mobility and extra pressure drop, and compare with previously reported results in the literature, in particular, the 3D BEM simulations of Ref.~\cite{barakat_shaqfeh_2018b} and the theoretical prediction of Ref.~\cite{barakat_shaqfeh_2018a} as confinement is approaching its critical value. Furthermore, we present several correlations and discuss their practical implications. A summary of the main findings and concluding remarks are presented in Sec.~\ref{Conclusions}. A lubrication theory, consolidated from previously reported works in the literature, is supplied in the appendix to provide scaling relations in support of the present numerical results.
  
\section{Governing equations: A fluid-structure interaction problem} \label{pb}

\subsection{Hydrodynamics}

\begin{figure}[!htbp]
  
  \centering {\includegraphics[scale=1.2]{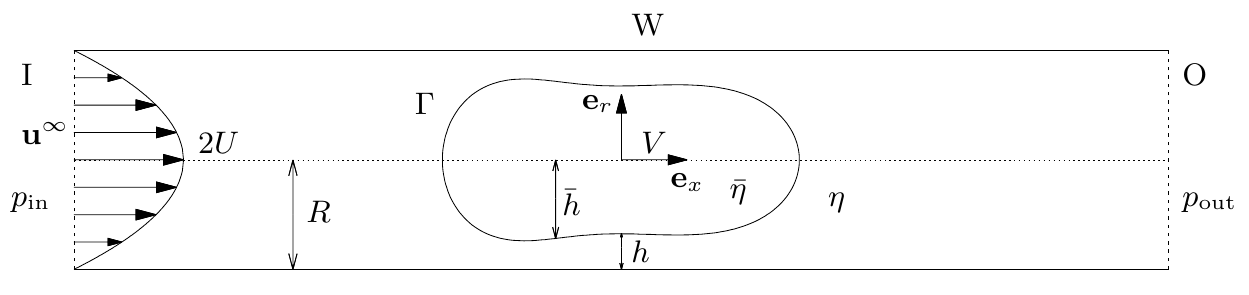}}
 \caption{Schematic illustration of a vesicle flowing along the centreline of a circular tube of radius $R$ in a pressure-driven flow. The system is rotationally symmetric about the $x$-axis. The boundaries of the control volume ($D$) are the inlet and outlet sections $I$ and $O$, the solid surface of the tube wall $W$, and the membrane/medium interface $\Gamma$, i.e., $\partial D \equiv I \cup O \cup W \cup \Gamma$. The vesicle shape and the gap size between the membrane and the tube wall are denoted by $\bar{h}$ and $h (\equiv R-\bar{h})$, respectively. The vesicle enclosed volume is denoted by $\Omega$.}
 \label{fig_plan}
\end{figure}

We consider a neutrally buoyant vesicle that is transported in a pressure-driven flow through a straight, circular tube of radius $R$. Fluid flows at an imposed, constant volumetric flow rate $Q$~($\equiv \pi R^2 U$, giving a mean bulk velocity $U$) driven by a pressure difference between inlet and outlet cross-sections. A schematic description of the problem is shown in Fig.~\ref{fig_plan} with axisymmetric cylindrical basis ($\mathbf{e}_x, \mathbf{e}_r, \mathbf{e}_{\phi}$). The suspending fluid and the fluid inside the vesicle membrane are incompressible and  Newtonian with dynamic viscosity $\eta$ and $\bar{\eta}$, respectively. We assume that the Reynolds numbers for both interior and exterior fluids are sufficiently small compared with unity, so that the inertial terms in the equations of motion may be neglected, the governing equations thereby reduce to the Stokes equations for creeping motion. Under these circumstances, experimental evidence~\cite{Europhysics_Vitkova_2004} shows (and we will assume) that the flow and vesicle shapes are axisymmetric and that a vesicle initially positioned at the tube axis will translate with a steady velocity $V\mathbf{e}_x$ (which is unknown {\it a priori}) along the axis of the tube.

In the absence of any vesicle, we obtain an unperturbed Poiseuille flow with a parabolic velocity profile:

\begin{subequations} \label{E:Poi}
  \begin{gather}
  \mathbf{u}^{\infty}(\mathbf{x}) = 2 U \left (1 - \frac{r^2}{R^2} \right ) \mathbf{e}_x, \qquad \mathbf{x} \in \mathrm{D},  \label{E:Poi1} \\
   \mbox{with} \qquad U = \frac{G^0 R^2}{8 \eta}, \quad  \Delta p^0 = \frac{8 \eta UL_w}{R^2}, \quad R^0_H= \frac{8 \eta L_w}{\pi R^4} , \label{E:Poi2} 
      \end{gather}
\end{subequations}
where $G^0$ is the negative (uniform) pressure gradient applied along a tube of length $L_w$, namely, $G^0=\Delta p^0/L_w$, $\Delta p^0$ is the pressure difference or pressure drop required for the Poiseuille flow, and $R^0_H$ is the hydraulic resistance ($\equiv \Delta p^0/Q$). 

The presence of a vesicle causes a change in the hydraulic resistance of the system (i.e., tube + vesicle): it increases, meaning that to maintain the same volumetric flow rate $ \pi R^2 U$, it is necessary to increase the pressure difference by a quantity $\Delta p^+$ called extra pressure drop. %The total pressure drop from the inlet to the outlet of the tube is found by adding the undisturbed pressure drop $\Delta p^0$ to the extra pressure drop $\Delta p^+$ due to  the vesicle. 
 Finding that extra pressure drop, together with the vesicle shape deformation and its mobility, is the essential part of the investigations in the present work. 

In the creeping-flow regime, the motion of the fluid inside and outside the vesicle is governed by the Stokes equations,

 \begin{equation}
 \bm{\nabla} \cdot \mathbf{u} = 0, \quad  \bm{\nabla} \cdot \bm{\sigma} = - \bm{\nabla} p + \eta \nabla^2  \mathbf{u} = \mathbf{0}, \qquad \mathbf{x} \in \mathrm{D}\setminus \Omega, \label{eq:STOKES} 
\end{equation}
where $\mathbf{u}$, $\bm{\sigma}$ and  $p$ denote the exterior fluid velocity,  stress tensor and  pressure, respectively. Similar equations hold for the interior fluid velocity $\bar{\mathbf{u}}$, stress tensor $\bar{\bm{\sigma}}$ and  $\bar{p}$ for $\mathbf{x} \in \Omega$. The fluid motion is coupled with the interface motion via the kinematic boundary condition,
\begin{equation}
 \mathbf{u}(\mathbf{x}) = \bar{\mathbf{u}}(\mathbf{x}) = \frac{\mathrm{d} \mathbf{x}}{\mathrm{d} t} = \mathbf{u}_\Gamma, \qquad \mathbf{x} \in \Gamma,
\label{eq:kinem}
\end{equation}
where $\mathbf{x}$ is the membrane position. The dynamic boundary condition at the interface establishes a \textit{nonlinear interaction} between the ambient flow and membrane mechanics, 
\begin{equation}
 \Delta \mathbf{f} + \mathbf{f}^m = \mathbf{0}, \qquad \mathbf{x} \in \Gamma,
\label{eq:dynam}
\end{equation}
wherein we assume the membrane is in quasi-static mechanical equilibrium; the membrane force density $\mathbf{f}^m$ balances the net traction $\Delta \mathbf{f}$~($\equiv (\bm{\sigma} - \bar{\bm{\sigma}}) \cdot \mathbf{n}$) exerted on the membrane by the bulk fluids. Here and throughout this paper, the unit normal vector of a surface $\mathbf{n}$ is pointing inward into the suspending fluid.

Additional boundary conditions for the velocity field include the no-slip condition at the tube wall,
\begin{equation}
\mathbf{u}(\mathbf{x}) = \mathbf{0}, \qquad \mathbf{x} \in W,
\label{eq:W}
\end{equation}
and vanishing far-field flow perturbation,
\begin{equation} 
 \mathbf{u}(\mathbf{x}) =   \mathbf{u}^{\infty}(\mathbf{x}), \qquad \mathbf{x} \in I \cup O .
 \label{eq:IO}
\end{equation}

The velocity of the vesicle center of mass is given by
\begin{equation} 
V =  \frac{1}{\Omega}  \int_\Omega \bar{u}_x \mathrm{d}^3\mathbf{x} = \frac{1}{\Omega} \int_\Gamma x (\mathbf{u}_\Gamma \cdot \mathbf{n})\mathrm{d}S(\mathbf{x}) .
 \label{eq:Vel}
\end{equation}
The enclosed volume 
\begin{equation} 
\Omega =   \int_\Omega \mathrm{d}^3\mathbf{x} = \frac{1}{3} \int_\Gamma (\mathbf{x} \cdot \mathbf{n})\mathrm{d}S(\mathbf{x})
 \label{eq:Vol}
 \end{equation}
 is fixed, as the vesicle membrane is considered to be impermeable, at least on typical experimental time scales. The axial coordinate of the vesicle center of mass is defined by
 \begin{equation}
\label{eq:mass }
x_G = \frac{1}{\Omega}\int_\Omega x \mathrm{d}^3\mathbf{x} =  \frac{1}{2\Omega}\int_\Gamma x^2 (\mathbf{n}\cdot \mathbf{e}_x) \mathrm{d}S(\mathbf{x}).
\end{equation}

\subsection{Membrane mechanics}
A biomembrane is invariably a lipid bilayer, which is classically described as a two-dimensional, incompressible fluid elastic. This means that there exist a surface tension and bending energy associated with the "out-of-the-plane" motions of the membrane. Its elastic energy due to the Helfrich energy functional~\cite{Helfrich_1973} is given by
\begin{equation}
E = \int_\Gamma \left [2\kappa H^2(\mathbf{x}) + \gamma(\mathbf{x}) \right ] \mathrm{d} S(\mathbf{x}), 
\label{eq:Helfrich}
\end{equation}
where $\kappa$ ($\sim 10^{-19}$~J) is the bending modulus, $H$ is the local mean curvature (with the convention that curvature is positive for a sphere), and $\gamma$ is the membrane tension, which is, in fact, identical  with the Lagrange multiplier used to enforce the surface incompressibility condition,
\begin{equation}
\bm{\nabla}_\mathrm{S} \cdot \mathbf{u}_{\Gamma} = 0, \qquad \mathbf{x} \in \Gamma,
\label{eq:surface}
\end{equation}
where $\bm{\nabla}_\mathrm{S}=(\mathbf{I} - \mathbf{n}\mathbf{n}) \cdot \bm{\nabla}$ is the surface gradient. 

The membrane force density $\mathbf{f}^m$, by the principle of virtual work, is the variational derivative of Eq.~(\ref{eq:Helfrich}) with respect to small deformations of the surface~\cite{Zhong_can_1989},
\begin{subequations} \label{E:fm}
  \begin{gather}
  \mathbf{f}^m = - \frac{\delta E}{\delta \mathbf{x}} = \mathbf{f}^b + \mathbf{f}^{\gamma}, \qquad \mathbf{x} \in \Gamma, \label{E:fm1} \\
\mbox{with} \quad \mathbf{f}^b = 2\kappa \left [ \Delta_\mathrm{S} H + 2 H(H^2 - K) \right ] \mathbf{n}, \quad \mathbf{f}^\gamma = -2\gamma H \mathbf{n} + \bm{\nabla}_\mathrm{S} \gamma,\label{E:fm2} 
 \end{gather}
\end{subequations}
where $\mathbf{f}^b$ denotes the bending surface force density, which is purely normal, $\mathbf{f}^\gamma$ is the tension surface force density, $K$ is the Gaussian curvature, and $ \Delta_\mathrm{S} H = \bm{\nabla}_\mathrm{S} \cdot \bm{\nabla}_\mathrm{S} H$ is the Laplace-Beltrami operator of the mean curvature, which contains the fourth derivative of the surface position, posing serious algorithmic and numerical challenges to compute the bending forces~\cite{Guckenberger_CPC_2016}.

\subsection{Dimensionless parameters}

The volume $\Omega$ and surface area $A$ of a vesicle remain constant and define a volume-based radius $R_0 \equiv (3\Omega/(4\pi))^{1/3}$ and an area-based radius $R_A\equiv (A/(4\pi))^{1/2}$, respectively. Together with the tube radius $R$, the system geometry is completely parametrized by two dimensionless  parameters which are independent of the flow conditions: the reduced volume $\nu$ (alternatively, the excess area $\Delta$) and the confinement $\lambda$,
\begin{subequations} \label{E:nu}
  \begin{gather}
\nu \equiv \frac{\frac{4}{3}\pi R_0^3}{\frac{4}{3}\pi R_A^3} = \left (\frac{R_0}{R_A} \right )^3 = 6 \sqrt{\pi} \Omega A^{-3/2}, \label{E:nu1} \\
\Delta \equiv 4 \pi \left ( \frac{4\pi R^2_A}{4\pi R^2_0} - 1 \right ) = 4 \pi \left ( \frac{1}{\nu^{2/3}} - 1 \right ),  \label{E:nu2} \\
\lambda \equiv \frac{R_0}{R}. \label{E:nu3}
\end{gather}
\end{subequations}
Here, we use the volumetric radius $R_0$ as the reference length. The reduced volume  ($0<\nu\leq1$) or the excess area ($\Delta \geq 0$), characterizing the ability for the vesicle to deform and change shape, is commonly used in the literature, and they are related to each other by~(\ref{E:nu2}). The confinement measures the size of the vesicle relative to the radius of the tube. Natural scales for velocities and time are the mean velocity $U$ of the ambient flow and  $R_0/U$, respectively. Pressure and hydrodynamic stress are scaled by the typical viscous stress $\eta U/R_0$, and membrane tension is scaled by $\eta U$. The relative importance of membrane bending force density and viscous traction in the balance of normal stress on the membrane (Eqs.~(\ref{eq:dynam}) and (\ref{E:fm2})) defines the bending-based capillary number $\mathrm{Ca_B}\equiv\eta U R^2_0/\kappa$. Finally, there is no viscosity contrast between the fluid inside and outside the vesicle as we are interested in the stationary axisymmetric shapes which do not depend upon the inner viscosity~\cite{SecombSkakak1986,Bruinsma1996}. Hence, the vesicle motion is determined by three independent dimensionless parameters: the reduced volume $\nu$, the confinement
 $\lambda$, and the (bending) capillary number $\mathrm{Ca_B}$. We note that while $\nu$ is a fixed quantity for a given vesicle, namely independent of which reference length is used, the other two parameters depend on that length. However, solutions under different scalings are easily converted from one to another in terms of $\nu$ and $\lambda$. For example, the surface area-based confinement $\lambda_A \equiv R_A/R = \lambda /\nu^{1/3}$, and the tube's radius-based capillary number $\mathrm{Ca_R} \equiv \eta U R^2/\kappa = \mathrm{Ca_B}/\lambda^2$. Occasionally, we nondimensionalize physical quantities using the scalings of cited references in order to facilitate the comparison with those results. 
 
 %Using $R_0$ as the unit length will make it easier to compute vesicle shapes, especially under high confinements. It's also more straightforward to interpret  

\section{Boundary element method simulation} \label{Num}
The fluid-cell membrane interaction problem described in Sec.~\ref{pb} is solved using an axisymmetric boundary element method (BEM)~\cite{Trozzo_JCP_2015}, which is based on the numerical method for 3D model~\cite{Boedec_JCP_2011}, therefore, only the complementary information is provided below.

First, in view of the linearity of the Stokes equations, we decompose the total velocity field around the vesicle into an undisturbed component $\mathbf{u}^{\infty}$ and a disturbance component $\mathbf{u^+}$ due to the presence of the vesicle, namely, $\mathbf{u} = \mathbf{u}^{\infty} + \mathbf{u^+} $. The disturbance velocity at a point $\mathbf{x}_0$ that lies inside the control volume $D$ or on its boundaries $\partial D$ can be represented as a boundary integral equation~\cite{Pozrikidis_1992,Pozrikidis_2005},

\begin{equation}
\mathbf{u}^+(\mathbf{x}_0) = - \frac{1}{8\pi\eta} \int_{\partial D} \mathbf{G}(\mathbf{x}_0, \mathbf{x}) \cdot \mathbf{f}^+(\mathbf{x}) \mathrm{d} S(\mathbf{x}), 
\label{eq:BEM1}
\end{equation}
where $\mathbf{G}$ is the free-space Green's function, $\mathbf{f}^+\equiv \bm{\sigma}^+ \cdot \mathbf{n}$ is the disturbance surface traction. Since the perturbation flow in tube generated by a point-force distribution decays exponentially with distance from the vesicle~\cite{Liron_1978,Trozzo_JCP_2015}, if the inlet and outlet are sufficiently far from the vesicle, then the flow perturbation near the inlet and outlet sections virtually vanishes. Furthermore, if we consider axisymmetric flow configuration only, the surface integrals can be explicitly integrated in the azimuthal direction with $\mathrm{d}S = r \mathrm{d}\phi \mathrm{d}l$, where $\mathrm{d}l$ is the differential arc length of the trace of the boundary $\partial D$ in the $x$-$r$ azimuthal plane~\cite{Pozrikidis_1992,Trozzo_JCP_2015}.  Finally, we obtain a more specific form to Eq.~(\ref{eq:BEM1}), yielding the total velocity field as follows:

\begin{eqnarray}
u_\alpha(\mathbf{x}_0) & = & u_\alpha^\infty(\mathbf{x}_0) + \frac{1}{8\pi\eta} \left[ \int_\Gamma M_{\alpha\beta}(\mathbf{x}_0, \mathbf{x}) f^m_\beta (\mathbf{x}) \mathrm{d} l(\mathbf{x}) 
- \int_W M_{\alpha\beta}(\mathbf{x}_0, \mathbf{x}) f^w_\beta (\mathbf{x}) \mathrm{d} l(\mathbf{x})  \right. \nonumber\\
& &\left. + \mbox{ } p^{+}_{\mathrm{in}} \int_I M_{\alpha x}(\mathbf{x}_0, \mathbf{x}) \mathrm{d} l(\mathbf{x}) \right] , 
\label{eq:BEM2}
\end{eqnarray}
where  the Greek subscripts $\alpha$ and $\beta$ are either $x$ or $r$, representing the axial and radial components respectively. Here, $\mathbf{x} = x\mathbf{e}_x + r\mathbf{e}_r$, $\mathbf{M}$ is the free-surface axisymmetric Green's function~\cite{Pozrikidis_1992,Trozzo_JCP_2015}, $\mathbf{f}^w (\equiv f^w_x\mathbf{e}_x + f^w_r\mathbf{e}_r)$ stands for the disturbance stress distribution at the tube wall with the shear stress $f^w_x$ and the normal stress $f^w_r$($=-p^+, \mathbf{x} \in W$), and $p^{+}_{\mathrm{in}}$ the disturbance pressure over the inlet while setting, without loss of generality, $p^{+}_{\mathrm{out}}=0$ for the disturbance pressure over the outlet.

Application of Eqs.~(\ref{eq:W}) and~(\ref{eq:BEM2}) leads to an additional integral boundary equation that allows the calculation of the disturbance wall stress $\mathbf{f}^w$,

\begin{eqnarray}
 \int_W M_{\alpha\beta}(\mathbf{x}_0, \mathbf{x}) f^w_\beta (\mathbf{x}) \mathrm{d} l(\mathbf{x}) & = &  \int_\Gamma M_{\alpha\beta}(\mathbf{x}_0, \mathbf{x}) f^m_\beta (\mathbf{x}) \mathrm{d} l(\mathbf{x}) \nonumber \\
 & & + \mbox{ } p^{+}_{\mathrm{in}} \int_I M_{\alpha x}(\mathbf{x}_0, \mathbf{x}) \mathrm{d} l(\mathbf{x}), \qquad \mathbf{x}_0 \in W.
\label{eq:BEM4}
\end{eqnarray}

The extra pressure drop can be obtained using the reciprocal theorem~\cite{Pozrikidis_2005} of the Stokes flow; it is expressed in terms of the membrane load and the ambient velocity field,
\begin{eqnarray}
\Delta p^+ \equiv p^{+}_{\mathrm{in}} - p^{+}_{\mathrm{out}} & = & -\frac{1}{Q} \int_\Gamma \mathbf{f}^m (\mathbf{x}) \cdot \mathbf{u}^{\infty} (\mathbf{x}) \mathrm{d} S(\mathbf{x}) \nonumber \\
& = &  -\frac{4}{R^2}\int_\Gamma r \left (1-\frac{r^2}{R^2}\right ) f^m_x (\mathbf{x}) \mathrm{d} l (\mathbf{x}).
\label{eq:deltap1}
\end{eqnarray}

The membrane is discretized by $N^m$ piecewise linear 2D elements, consisting of a collection of points of $\{\mathbf{x}_n(t), n \in 0\ldots N^m\}$. The points are distributed according to the magnitude of the membrane's mean curvature $H$, thereby allowing the local mesh refinement in high-curvature regions. This is important given large deformations of vesicles involved in flows. The mesh points at the tube wall, composed of $N^w$ linear elements, are uniformly distributed. The differential surface operators, which are involved in the calculation of surface incompressibility~(\ref{eq:surface}) and bending forces~(\ref{E:fm2}), are computed on each element of the membrane with a parametrization $(\Gamma, \phi)$ of the surface.

Starting from some initial configuration of the vesicle shape, the preceding three integral equations, together with the membrane incompressibility condition~(\ref{eq:surface}), allow the computation of $\Delta p^+$, $\mathbf{f}^w$, the interfacial velocity $\mathbf{u}_\Gamma$ and the membrane tension $\gamma$ at each time step via the boundary element method, as described in~\cite{Trozzo_JCP_2015,Boedec_JCP_2011}. The vesicle translational velocity $V$ is computed from~(\ref{eq:Vel}). 

The vesicle interface is advected according to 
\begin{equation}
\label{eq:evolu}
\frac{\mathrm{d}\mathbf{x}(t)}{\mathrm{d}t}= u_n(\mathbf{x})\mathbf{n}(\mathbf{x}),
\end{equation}
where $\mathbf{x}$ is an interface node and $u_n=\mathbf{u}_\Gamma \cdot \mathbf{n}$ is given by~(\ref{eq:BEM2}). This means the movements of the bilayers in the normal and tangential directions are treated differently, namely in Lagrangian fashion for the former and with a Eulerian description for the latter. Indeed, the tangential movement of nodes, which does not change the membrane shape, offers the possibility of a redistribution of nodes--remeshing along the membrane.  At each time we employ a keeping-the center-of-mass strategy that the vesicle is re-centered at the origin by subtracting the vesicle center of mass $(x_G, 0)$ from the membrane position. Since this process, which is equivalent to replacing $u_n$ by $u'_n=u_n - \mathbf{V}\cdot \mathbf{n}$~in~(\ref{eq:evolu}), does not modify the stress field, the vesicle shape remains unchanged. Equation~(\ref{eq:evolu}) is solved numerically by a semi-implicit time stepping scheme~\cite{Boedec_JCP_2011} in which the bending forces are computed at the advected, new position of the membrane and, therefore improving long term stability of the algorithm. A steady state is obtained when the maximum absolute normal velocity $|u'_n|/U$ is less than a chosen tolerance ($\sim 10^{-5}$).  The number of elements  $N^m$ varies from 130 to 500, and $N^w$ from 300 to 1500, depending upon the reduced volume $\nu$ and the confinement $\lambda$. The tube has a total length of $L_w = O(10R_0)$, so the outlet and inlet sections are located at a distance $x=\pm L_w/2$ from the vesicle center of mass. The dimensionless time step $U \Delta t/R_0$ ($\sim 5\times10^{-3}-5\times10^{-5}$) decreases as $\nu$ decreases and as either $N^m$ or $\lambda$ increases. Since both the volume and surface area of the vesicle are conserved, the change in the enclosed volume and surface area during simulations provides an indication of the accuracy of the computations. The relative volume and surface area variations were found to be $\sim 0.01\%-0.1\%$ over a typical full simulation ($\sim 10^4-10^5$ time steps).
 
 \begin{table}[!h]
   \caption{\label{tab:comp}Comparison of droplet relative velocity $V/U$ and dimensionless extra pressure drop $\Delta p^+/(\eta U/R)$ as a function of the capillary number $\mathrm{Ca}=\eta U/\gamma$ for $\lambda=0.8$ to those reported in Ref.~\cite{Lac_JFM2009} for droplets at unit viscosity contrast.}
\begin{ruledtabular}
\begin{tabular}{lcccc}
   &  \multicolumn{2}{c} {$\hspace{0.7cm}V/U$ } &  \multicolumn{2}{c} {$\hspace{0.8cm}\Delta p^+/(\eta U/R)$}\\ 
       Ca & Present work  & Ref.~\cite{Lac_JFM2009} & Present work & \textrm{Ref.~\cite{Lac_JFM2009}} \\  \hline
        0.05	 &  1.4218 & 1.42 &  2.6591 & 2.65  \\ 
      0.1 & 1.4543 &   1.45 & 2.2532 & 2.25  \\ 
   0.2 &  1.5311 & 1.53 & 1.4981 & 1.50    \\ 
   0.3 & 1.5999 & 1.60  & 1.0063   & 1.01 \\ 
     0.5 & 1.7021 & 1.70   & 0.4889  & 0.49 \\ 
         \end{tabular}
         \end{ruledtabular}
 \end{table}

While several studies have already been conducted in~\cite{Trozzo_JCP_2015} to validate the axisymmetric BEM code, there is a need to check the numerical procedure that calculates the hydrodynamical quantities like the relative velocity $V/U$ and extra pressure drop $\Delta p^+$. This procedure, which is equivalent to data post-processing, is independent of soft objects studied as long as the shape of the soft object and its interfacial velocity and stress are given (cf. Eqs.~(\ref{eq:Vel}) and (\ref{eq:deltap1})). A simple way of validating such a calculation seems to compare the well-known example of a (clean, surfactant-free) drop in tube flow, for which very highly accurate numerical computations are available in the literature (e.g., Ref.~\cite{Lac_JFM2009}). The  motion of the drop at unit viscosity contrast is characterized by the capillary number $\mathrm{Ca}=\eta U/\gamma$ with $\gamma$ constant surface tension. The axisymmetric BEM code has been accommodated to simulate drop dynamics in tube flow in the following ways: (i) we set bending force $\mathbf{f}^b=\mathbf{0}$ while $\mathbf{f}^\gamma=-2H\mathbf{n}/\mathrm{Ca}$ (cf. Eq.~(\ref{E:fm2}); (ii) surface-area incompressibility, i.e., Eq.~(\ref{eq:surface}) is no longer used for drop dynamics because of a prescribed surface tension (or inversely the capillary number); and (iii) there is no viscosity contrast between inner and external flows. The motion of the drop for given confinement $\lambda$ is determined solely by the capillary number $\mathrm{Ca}$. Our numerical results are compared with those reported in Ref.~\cite{Lac_JFM2009}. The comparison in Table I shows excellent agreement.
 
In Figs.~\ref{fig_shapes2}, \ref{fig_vel}, and \ref{fig_dpa}, we also compared the present axisymmetric results with the 3D BEM simulations of~\cite{barakat_shaqfeh_2018b}. Note that the results of~\cite{barakat_shaqfeh_2018b} (their figure 4) were previously validated against our prior study~\cite{Trozzo_JCP_2015}. All of these comparisons show very good agreement between the two simulations.

Additionally, we have checked in two ways whether a numerical discretization, in terms of the number of elements (of both $N^m$ and $N^w$), was fine enough to sufficiently resolve the drainage fluid of thin liquid film between the membrane and the tube wall: (i) can a stationary solution be achieved ? and, most importantly, (ii) is the steady velocity ratio $V/U$ larger than unity? An insufficient discretization was found to result in  significantly large viscous and confinement-dependent friction on the membrane and therefore a smaller velocity ratio $V/U$, which sometimes is much less than unity. Under high confinement, especially for highly deflated vesicles (i.e., $\nu \leq 0.7$) when $\lambda$ approaches its critical value, a substantially large number of elements, say $N^m= O(500)$ and $N^w=O(1500)$ for $\nu=0.6$, is needed to accurately resolve the vesicle shape and membrane traction. Our numerical experiments suggest an empirical relationship between the mesh size and the film thickness $h$:
\begin{equation}
\delta x^m \approx h, \,\,\, \delta x^w \approx \frac{1}{2} h ,
\end{equation}
where  $\delta x^m$ and $\delta x^w$ denote a typical mesh size (i.e., element length) on the membrane and at the tube wall, respectively, in the region of the liquid film.

\section{Results and discussion} \label{Res}

\subsection{Phase diagram of shapes and shape transition}

In aqueous solution, lipid vesicles exhibit a large variety of shapes and shape transformations~\cite{Lipowsky_1991,Seifert_PRA1991}, in particular, they can exhibit a biconcave shape typical of red blood cells. The equilibrium shape of a vesicle is determined by minimization of the Helfrich energy (\ref{eq:Helfrich}) of the membrane, resulting in different families of solutions with respect to the reduced volume $\nu$. The global minimum is for a prolate if $\nu \in [0.652:1]$, an oblate if $\nu \in [0.592:0.651]$, and a stomatocyte if $\nu \in [0.05:0.591]$. When confined in capillary tubes subject to a pressure-driven flow, however, vesicles assume complex shapes and behave in different ways due to the nonlinear interplay between bending elasticity,  hydrodynamic stresses, and confinement. For the axisymmetric case being considered in this study, there are two commonly steady-state shapes which are classified as bullet-like and parachute-like shapes, the latter being characterized by a concave (negative curvature) rear part.

\begin{figure}[!htbp]
  \centering
  \includegraphics[scale=0.8]{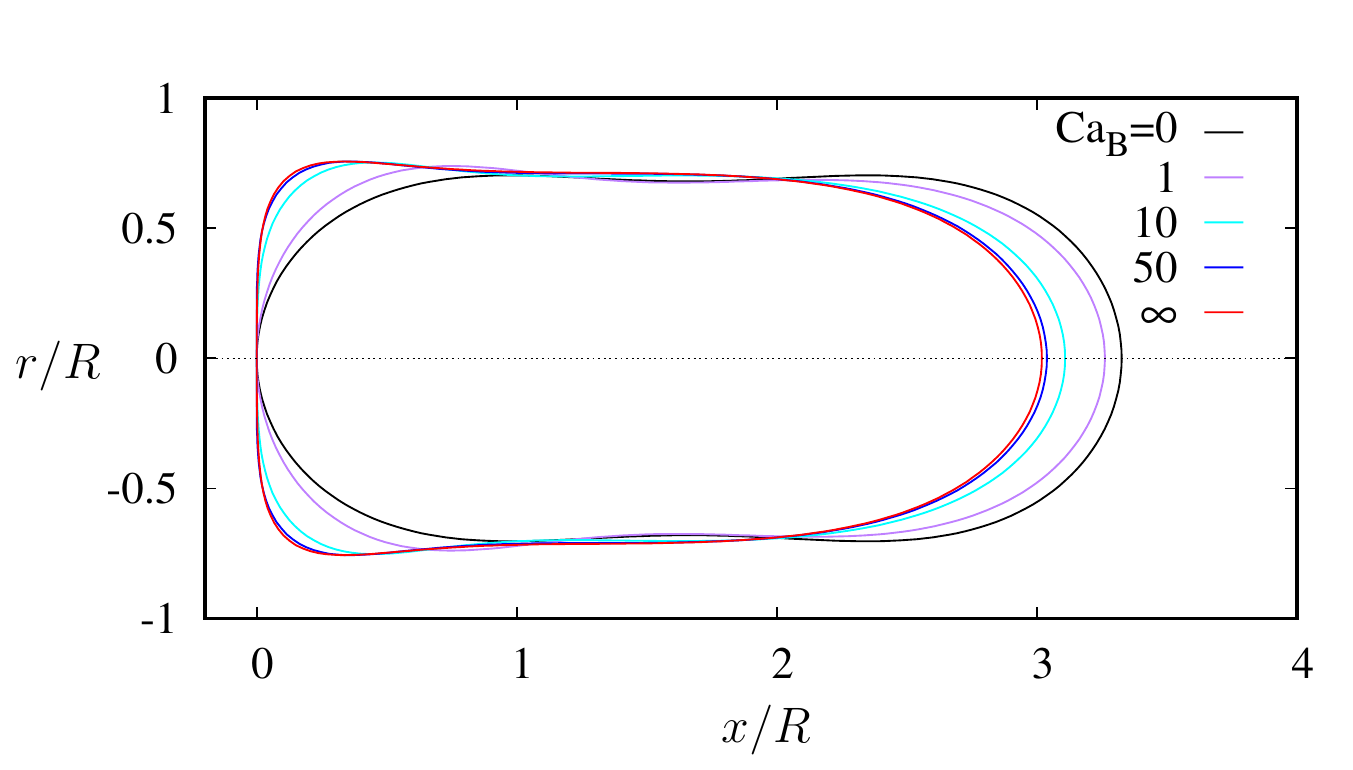}
 \caption{Steady-state vesicle profile as a function of the bending capillary number $\mathrm{Ca_B}$ for $\nu=0.84$ and $\lambda=1$.}
 \label{fig_shapes1}
\end{figure}

\begin{figure}[!htbp]
  \centering
  \includegraphics[width=14cm]{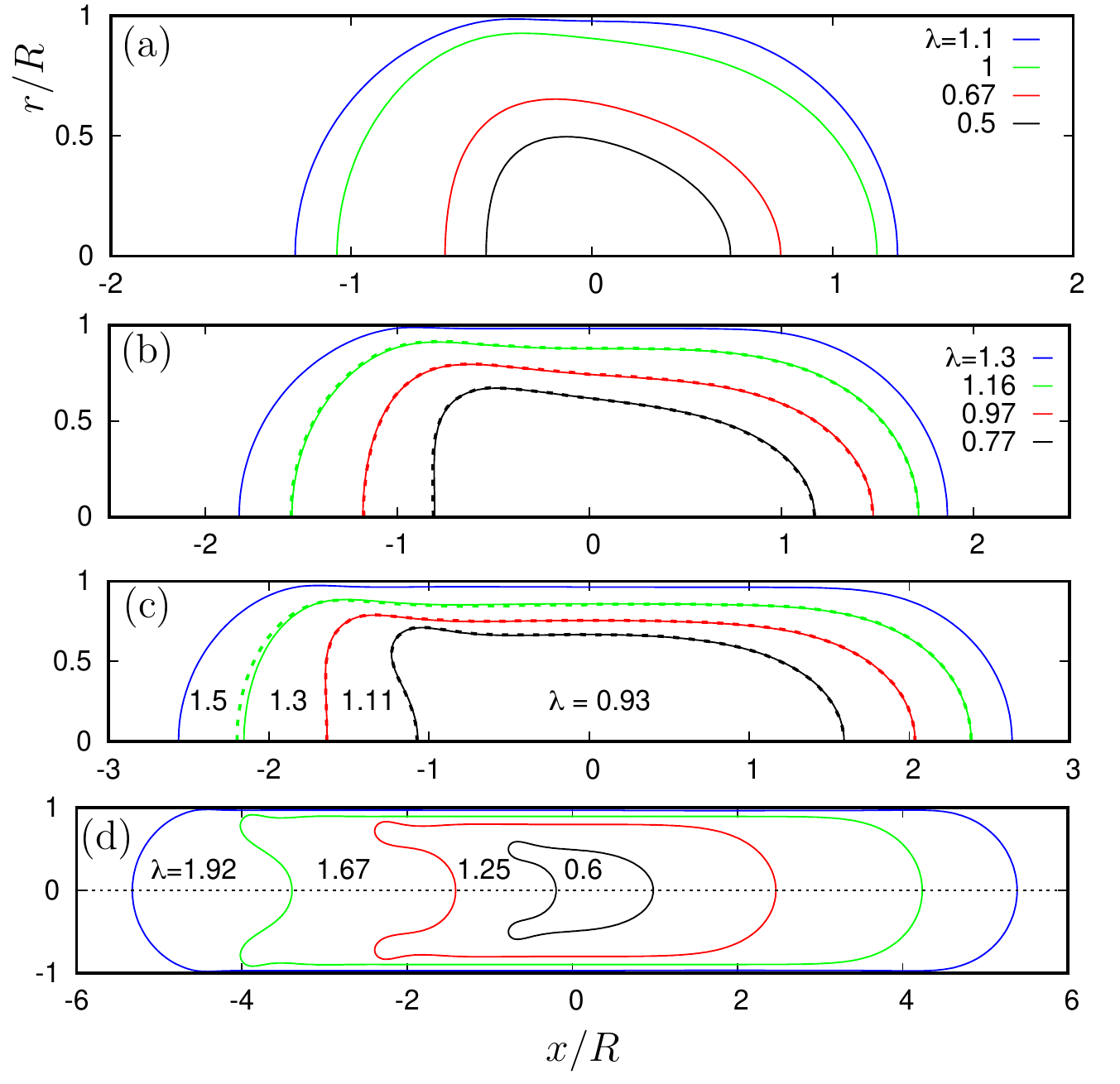} 
 \caption{Steady-state vesicle shapes (solid lines) as a function of the confinement $\lambda$ for a wide range of reduced volumes $\nu$ ($\mathrm{Ca_B=50)}$: (a) $\nu=0.98$, (b) $\nu=0.9$, (c) $\nu=0.8$, and (d) $\nu=0.6$. The dashed lines in (b) and (c) represent the 3D BEM simulations of~\cite{barakat_shaqfeh_2018b}, figure 9(b) ($\nu=0.9$, $\lambda_\mathrm{A} = 1.2$, 1, 0.8.) and figure 9(c) ($\nu=0.8$, $\lambda_\mathrm{A} = 1.4$, 1.2, 1.), respectively.}
 \label{fig_shapes2}
\end{figure}

\begin{figure}[!htbp]
  \centering
  \includegraphics[scale=0.8]{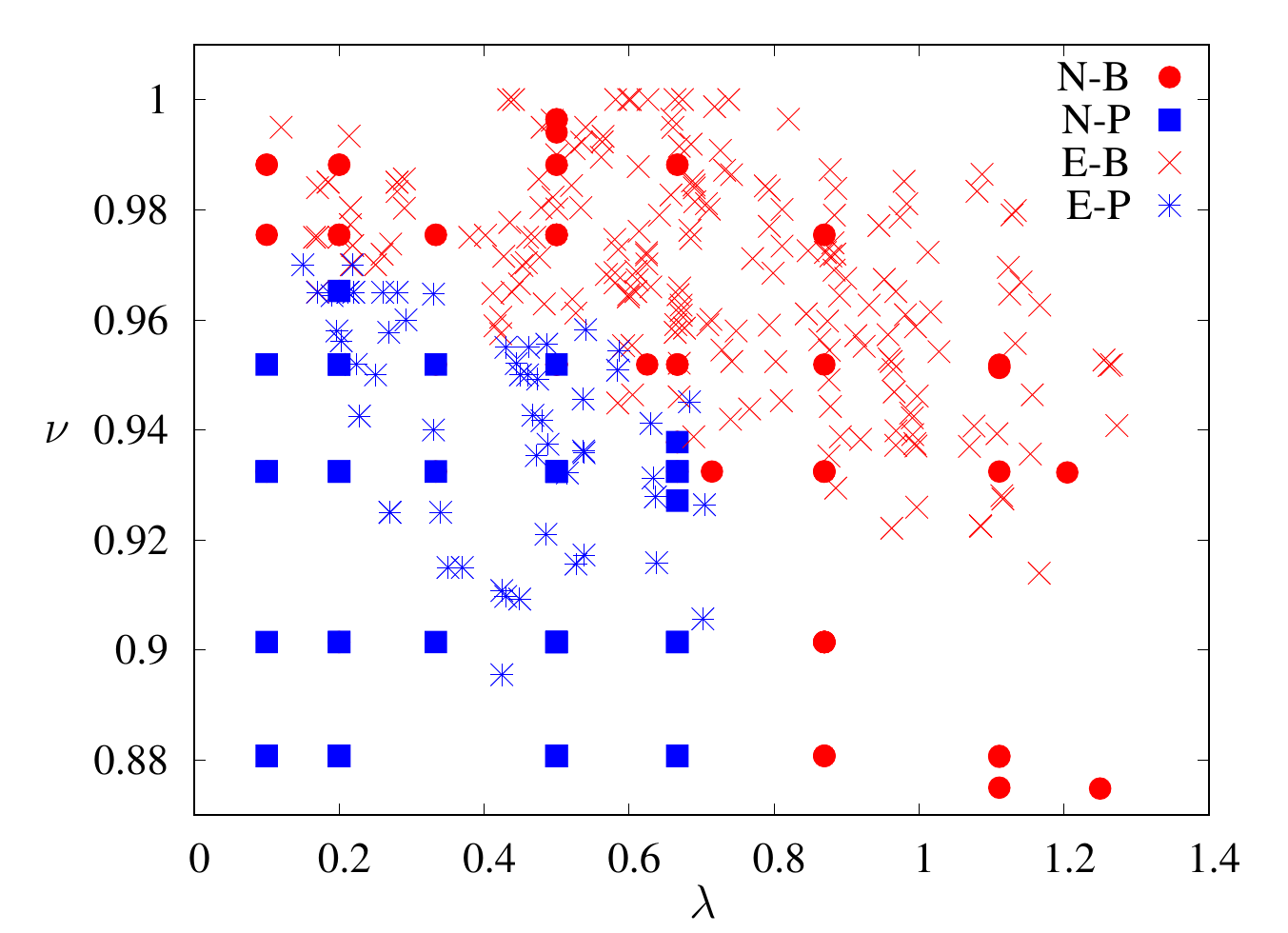}
 \caption{Parachute-bullet phase diagram in the ($\nu$, $\lambda$) space. Filled symbols denote the present numerical results (\textcolor{red}{\Large{$\bullet$}} for bullet-like shape, \textcolor{blue}{\small{$\blacksquare$}} for parachute-like shape), crosses represent experimental data of Ref.~\cite{Coupier_PhysRevLett108_2012} (\textcolor{red}{\large{$\times$}} for bullet-like shape, \textcolor{blue}{\Large{$\ast$}} for parachute-like shape). Capillary number varies over four orders of magnitude ($10^1 < \mathrm{Ca_B} < 10^4$).} \label{fig_B-P-phase-diagram}
\end{figure}

Figure~\ref{fig_shapes1} shows steady-state shapes for a vesicle of $\nu=0.84$ confined in a tube flow with unity radius ratio (i.e., $\lambda=1$) for increasing flow rates characterized by the bending capillary number $\mathrm{Ca_B}$. Two limiting cases are clearly illustrated in Fig.~\ref{fig_shapes1}. One is no flow ($\mathrm{Ca_B}=0$), in which the equilibrium shape -- symmetric between the front and the rear -- is determined solely by the minimization of the bending energy. The other is $\mathrm{Ca_B} \to \infty$, which corresponds to an infinitely small bending resistance. For this particular combination of parameter groups, the rear part of the vesicle becomes almost flat (i.e., zero curvature). This result is interesting because, for a given vesicle in a tube, its steady-state shape lies between these two limiting profiles. Another noticeable feature is that when $\mathrm{Ca_B} \ge 50 $, the shape is virtually independent of $\mathrm{Ca_B}$. This property allows us to fix $\mathrm{Ca_B} =50 $ while studying the motion and deformation of a confined vesicle at high flow rates (i.e., $U/(2R) > 50~\mathrm{s}^{-1}$)~\cite{SecombSkakak1986,Pries_1992}. A zero-bending elasticity or equivalently an infinity $\mathrm{Ca_B}$ is, however, not permitted because of small radii of curvature occurring at the trailing edge, especially for highly deflated vesicles.

Steady-state vesicle shapes for the reduced volume ranging from 0.98 to 0.6 are shown in Fig.~\ref{fig_shapes2} at different degrees of confinement, along with the numerical results of~\cite{barakat_shaqfeh_2018b}. The comparison in Fig.~\ref{fig_shapes2} shows almost perfect agreement between the two simulations, with a slight discrepancy only for $\nu=0.8$ and $\lambda=1.3$. A near spherical vesicle (i.e., $\nu=0.98$) always exhibits a bullet-like shape, whereas shapes undergo a transition from parachute to bullet as the confinement increases (i.e., large $\lambda$). The shapes with $\nu=0.6$, which are particularly relevant to red blood cells, mark a transition starting from a bell shape and ending in a sphero-cylinder. Clearly, increasing $\lambda$ increases the length of the vesicle but reduces the size of the gap between the vesicles and the tube. At high confinement, the vesicles tend to attain a  sphero-cylinder consisting of a long cylindrical main body and two hemispherical endcaps.

\begin{figure}[!htbp]
  \centering
 \includegraphics[scale=0.8]{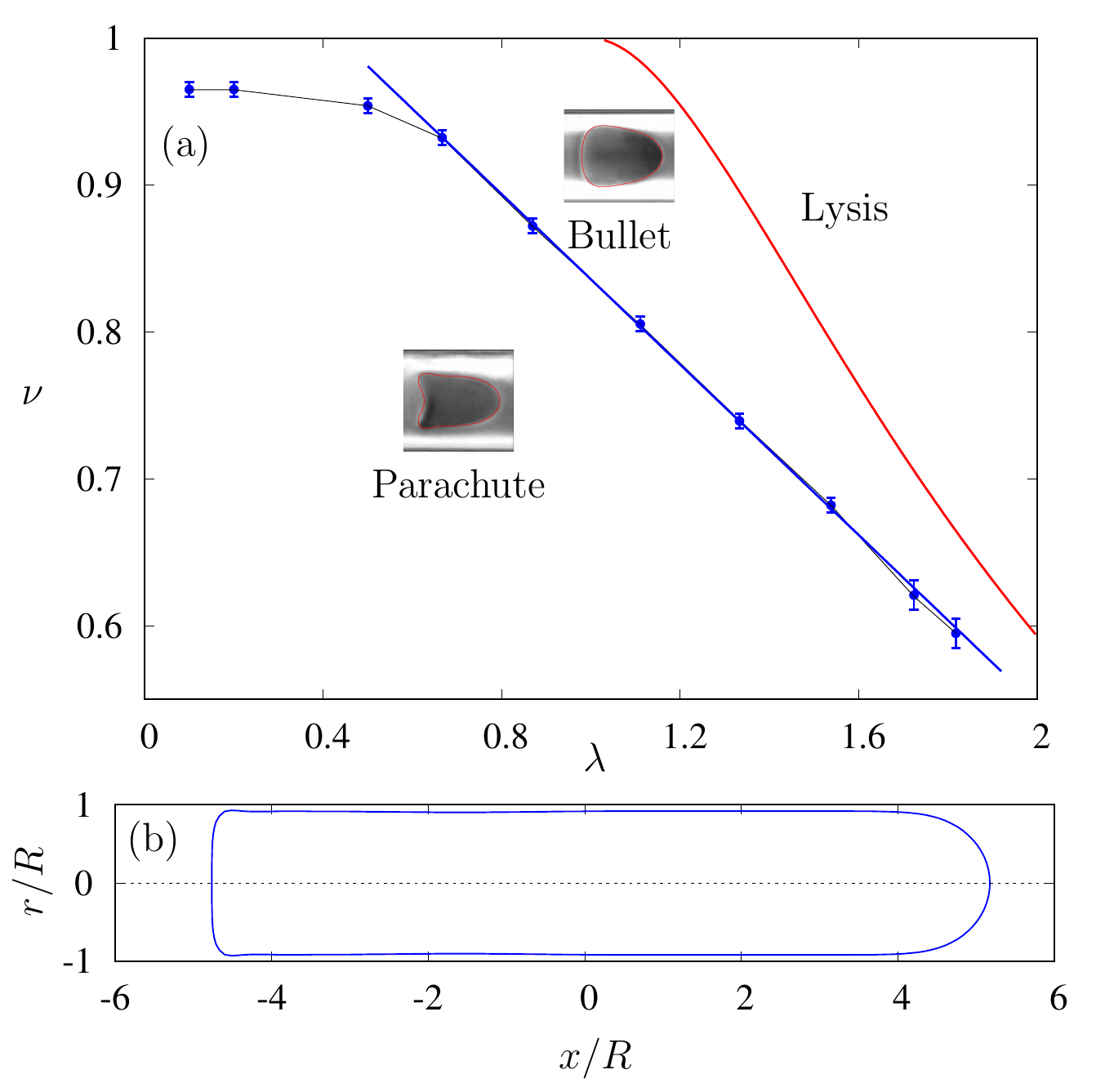}
 \caption{(a) Shape transition line (blue) from parachute (low limit in error bars) to bullet (up limit in error bars) in the ($\nu$, $\lambda$) space ($\mathrm{Ca_B}=50$). Vesicles are flowing from left to right. The two typical vesicles are characterized, respectively,  by $\nu = 0.95$, $\lambda = 0.8$ and $\mathrm{Ca_B}=5$ for the bullet-like shape, and $\nu = 0.83$, $\lambda = 0.67$ and $\mathrm{Ca_B}=15$ for the parachute-like shape. Insets show the comparison of the computed shapes (red line) with the reported ones in an experimental study \cite{Europhysics_Vitkova_2004}.  The rupture line (red) represents the critical confinement $ \lambda_c$~\eqref{eq:critic2}. Membrane lysis occurs when $\lambda > \lambda_c$. (b) The shape at the parachute-bullet transition  of a vesicle with the same typical deflation of red blood cells ($\nu =  0.6$) flowing through a narrow capillary tube ($\lambda \simeq 1.8$).}
	\label{fig_transition}
\end{figure}

 While the vesicle shapes in Fig.~\ref{fig_shapes2} have been reported elsewhere, e.g., in Refs.~\cite{SecombSkakak1986,Trozzo_JCP_2015,barakat_shaqfeh_2018b,Pozrikidis_2005b,Zhao_JCP_2010},
the new results are the phase diagram and comparison to the square-duct experiments of~\cite{Coupier_PhysRevLett108_2012}. As shown in Fig.~\ref{fig_B-P-phase-diagram}, increasing $\lambda$ makes the transition shifted towards lower values of reduced volume, which means a bullet-like shape is favored at high confinement. The present circular-tube simulations are in good agreement with the experimental observations, suggesting the geometry of the channel might not affect too much the parachute-bullet transition. Interestingly, there is a clear separation between the bullet region and parachute region in the ($\nu$, $\lambda$) space, as revealed in Fig.~\ref{fig_transition}(a). Bullet-like and parachute-like shapes are identified according to the curvature of the rear part of vesicles; a flat rear  marks as a transition point  in the ($\nu$, $\lambda$) space. Remarkably, the transition point decreases almost linearly with increasing confinement when $\lambda \ge 0.6$. A linear fitting to the transition points gives a correlation 
\begin{equation}
\nu_\mathrm{T}  = 1.126 -0.29 \lambda_\mathrm, \qquad \mbox{for} \quad  \nu \leq 0.93,
\label{eq:transition}
\end{equation}
which are obtained at $\mathrm{Ca_B}=50$. A careful examination of numerical results suggests this relationship is valid even for a small bending capillary number since steady shapes are virtually independent of $\mathrm{Ca_B}$ under high confinement. The transition from a parachute shape to a bullet one for a vesicle having the same reduced volume of red blood cells takes place at very high confinement, i.e., $\lambda_\mathrm{T} (\nu=0.6) \simeq 1.8$. The shape at the transition point, as shown in Fig.~\ref{fig_transition}(b), consists of a long cylindrical body and a front endcap, leaving a narrow vesicle-wall gap of $\sim 7\%$ of the tube's radius.

It is noticed that the parachute and bullet shapes can coexist around the phase boundary, the question then arises as to whether the boundary depends on the initial shape (parachute or bullet). In a previous study~\cite{Trozzo_JCP_2015}, it is confirmed that for large enough capillary numbers (i.e., $\mathrm{Ca_B} \ge 50$) the initial shape (prolate or oblate) does not change the steady vesicle shape though it does affect its dynamics during the transition. Here, we have also checked that the transition line does not depend on the initial shape (parachute or bullet).

The critical confinement $\lambda_c$~($\equiv 2R_0/d_c$), which corresponds to a lower limit $d_c$ to the diameter of the tubes through which a vesicle may pass intact, is calculated by assuming that the two hemispherical endcaps are tightly fitting the tube cross-section; its relation with the reduced volume $\nu$ is given by a cubic equation~\cite{Canham_1968,barakat_shaqfeh_2018a}
\begin{equation}
\label{eq:critic}
2\lambda_c^3 -3 \nu^{-2/3}\lambda_c^2 + 1 =0.
\end{equation}
One can solve~\eqref{eq:critic} for a critical $\nu_c$
\begin{equation}
\nu_c = \left ( \frac{3\lambda^2}{1+2\lambda^3}\right)^{3/2}.
\label{eq:critic2}
\end{equation}
We plot this equation in Fig.~\ref{fig_transition}(a), which represents the rupture line. When $\lambda > \lambda_c$, the vesicle cannot pass through the tube without rupturing its membrane.

This critical confinement defines an upper limit $\ell_c$ to the reduced vesicle length $\ell$ ($\equiv L/R$)
\begin{equation}
\ell_c \equiv \frac{2L_c}{d_c} = 2 \nu^{-2/3}\lambda_c^2.
\label{eq:lc}
\end{equation}
Assuming red blood cells have a typical volume $\Omega=90$~$\mu$m$^3$, which gives a volumetric radius $R_0 \simeq 2.8~\mu$m, one then obtains a critical cylindrical diameter of normal human erythrocyte $d_c=2 R_0/\lambda_c \simeq 2.8~\mu$m ($\lambda_c \simeq 1.98$ for $\nu=0.6$) and the maximum length of cells $L_c  \simeq  15.4~\mu$m ($\ell_c \simeq 11$). This means that a normal human erythrocyte can squeeze through capillaries that are smaller than half the diameter of a red blood cell ($\simeq 8~\mu$m).

\subsection{Lubrication film thickness}

\begin{figure}[!htbp]
  \centering
 \includegraphics[scale=0.8]{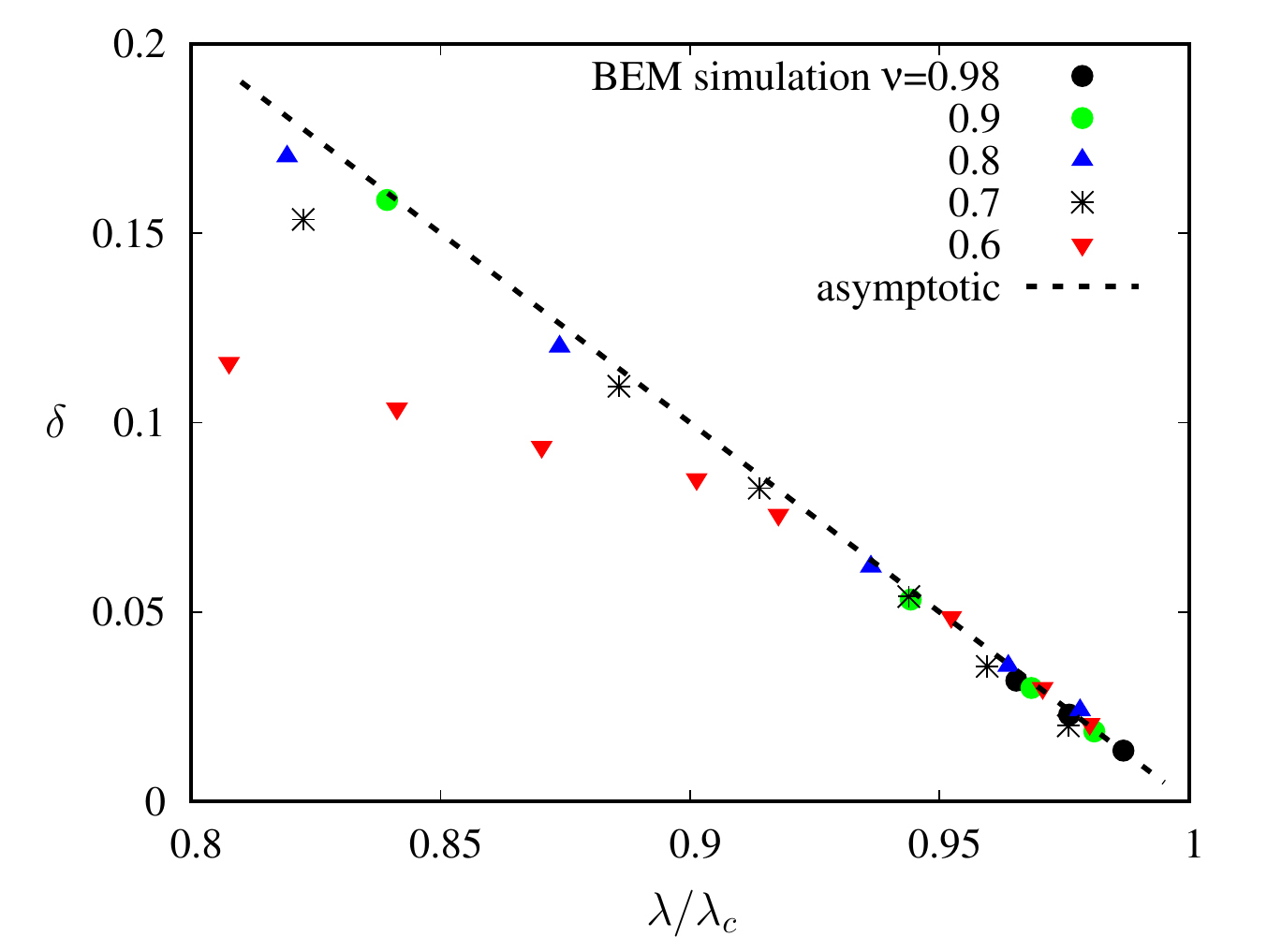}
 \caption{The dimensionless thickness of the lubrication layer $\delta$ plotted as a function of the reduced radius ratio $\lambda / \lambda_c$ for a wide range of reduced volumes $\nu$ ($\mathrm{Ca_B}=50$), together with the asymptotic prediction (\ref{eq:sfilm2}).}
\label{fig_film}
\end{figure}

As shown in Fig.~\ref{fig_shapes2}, under high confinement, a liquid film of nearly uniform thickness is formed between the front and rear endcaps. 
Let $h$ denote a typical film thickness of the gap (evaluated at the vesicle's midplane) separating the vesicle membrane and tube wall, the narrow-gap theory in the appendix yields an asymptotic behavior of the clearance parameter $\delta (\equiv  h/R)$ in terms of a small perturbation parameter $(1 - \lambda/\lambda_c) << 1$,
\begin{equation}
 \delta = 1 - \lambda/\lambda_c +  O\left [(1 - \lambda/\lambda_c)^2 \right].
 \label{eq:sfilm2}
\end{equation}
Numerical results of $\delta$ for $\nu$ ranging from 0.98 to 0.6 are plotted in Fig.~\ref{fig_film} and compared with its asymptotic behavior given by (\ref{eq:sfilm2}). Despite a wide range of the reduced volumes being considered, Fig.~\ref{fig_film}  makes it clear that when $1- \lambda/\lambda_c < 0.1$ -- namely in the small-gap regime, the numerical results approach the prediction (\ref{eq:sfilm2}); minimal thickness are about 2--$5\%$ of the tube's radius when $\lambda/\lambda_c \simeq 0.98$.

\begin{figure}[!htbp]
  \centering
 \includegraphics[scale=0.8]{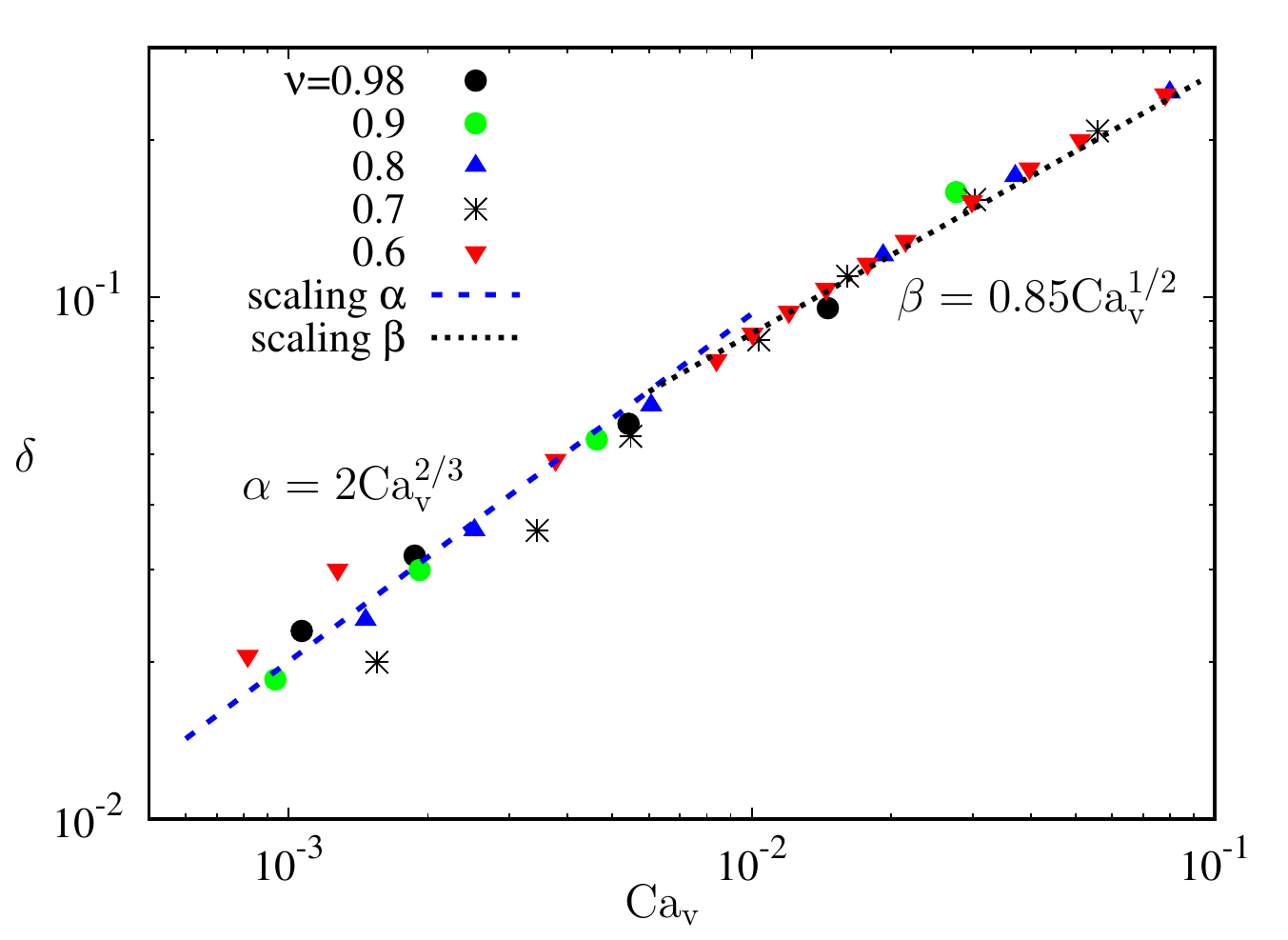}
 \caption{The dimensionless thickness of the lubrication layer $\delta$ plotted as a function of the capillary number $\mathrm{Ca_v}=\eta V/\gamma_\mathrm{F}$ for a wide range of reduced volumes $\nu$, together with two scalings.}
	\label{fig_film_ca}
\end{figure}

As derived in the appendix, the film thickness is controlled by the membrane tension $\gamma$ in the lubrication layer. The clearance size is found to be proportional to the 2/3 power of the capillary number $\mathrm{Ca_v}$ 
\begin{equation}
\label{eq:sfilm4}
\delta  \simeq c_0 \mathrm{Ca^{2/3}_v},
\end{equation}
where the vesicle tension-mobility-based capillary number $\mathrm{Ca_v} = \eta V/\gamma_F$, $\gamma_F$ is the membrane's frontal tension, also the highest tension in the membrane. We note that the numerical prefactor $c_0$ differs slightly in the literature; $c_0 \simeq 2.123$ in Ref.~\cite{SecombSkakak1986}  while $c_0 \simeq 2.05$ in Ref.~\cite{Bruinsma1996}.  A fitting to the present numerical results yields $c_0\simeq 2$ for $\mathrm{Ca_v}<10^{-2}$, as shown in Fig.~\ref{fig_film_ca}. In addition, Fig.~\ref{fig_film_ca} reveals a characteristic change in power scaling from the 2/3 power regime for small $\mathrm{Ca_v}$ to the 1/2 power regime for large $\mathrm{Ca_v}$. The separation of the two regimes occurs at $\mathrm{Ca_v} \simeq 6 \times 10^{-3}$. Therefore, our numerical results support one of the findings of Ref.~\cite{Bruinsma1996} that the thickness of the lubrication layer, at high flow rates,  is independent of the bending energy and is determined solely by the membrane tension. It should be emphasized that the 2/3 power law regime found in the case of a long bubble in tubes~\cite{bretherton_1961} stems from the different underlying mechanisms compared to vesicles; a stress-free surface for the former while a "no-slip" hydrodynamic boundary condition for vesicles. It is also important to point out that the vesicle tension-mobility-based capillary number $\mathrm{Ca_v}$, which is based on an immeasurable tension,  is not a controllable parameter unlike in the case for droplets and bubbles, where the surface tension is a material property.

The mechanical tension of a membrane is identical with the Lagrange multiplier tension $\gamma$  that enforces a certain, fixed membrane area \cite{Lipowsky_2014}. For the lipid bilayers, the rupture tension, which represents the largest mechanical tension that can be applied to the membrane, is of the order of a few mN/m. It is, therefore, interesting to examine whether our BEM simulations are indeed able to predict a mechanical tension approaching that limit.  For $\nu=0.6$ typical of red blood cells and under the maximum possible confinement that we have reached, i.e., $\lambda= 1.94$ ($\lambda_c \simeq 1.98$), the maximum dimensionless tension is found to be $\gamma_{\mathrm{max}} \simeq 1.2\times10^3$, which gives rise to a dimensional mechanical tension about 1.5 mN/m. Here the tension is scaled by $\gamma_\mathrm{ref}=\eta U = \kappa \mathrm{Ca_B} /R_0^2 = 1.28 \,\mu$N/m, with $\mathrm{Ca_B}=50$, $\kappa = 2 \times 10^{-19}\,$J and $R_0 = 2.8\, \mu$m. The predicted mechanical tension at the proximity of the maximum confinement is actually of the order of the rapture tension.

\begin{figure}[!htbp]
  \centering
 \includegraphics[scale=0.8]{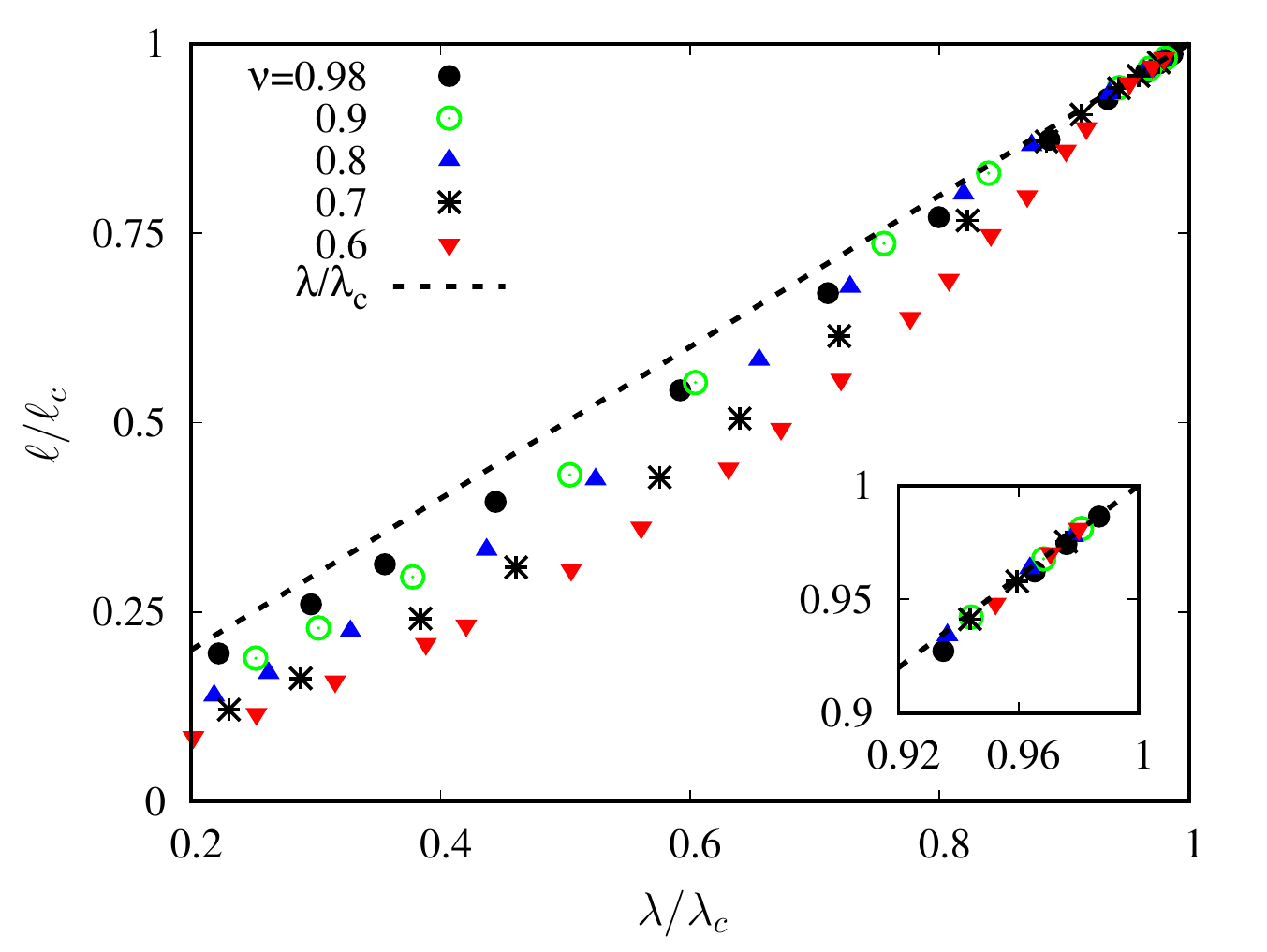}
 \caption{Normalized vesicle length $\ell/\ell_c$ versus the reduced radius ratio $\lambda/\lambda_c$ for a wide range of reduced volumes $\nu$ ($\mathrm{Ca_B}=50$), together with  the linear scaling $\ell/ \ell_c  =\lambda/\lambda_c$ (Eq. (\ref{eq:sL})).}
	\label{fig_length}
\end{figure}

The reduced vesicle length $\ell$ is an important geometric parameter that characterizes the vesicle deformation when flowing in a tube. It is also an easily accessible parameter in tube-flow experiments. In addition, it was found that there is an upper limit to the reduced vesicle length, given by Eq.~(\ref{eq:lc}).  Figure~\ref{fig_length} shows the normalized vesicle length $\ell/\ell_c$ as a function of the reduced radius ratio $\lambda/\lambda_c$ for a wide range of reduced volumes $\nu$. It is quite clear that the simulated vesicle lengths tend towards its limiting behavior as $\lambda \to \lambda_c$. The correlation   
\begin{equation}
 \ell = \ell_c \frac{\lambda}{\lambda_c} = 2\nu^{-2/3}\lambda_c^2 \frac{\lambda}{\lambda_c},
 \label{eq:sL}
\end{equation}
suggests an estimated vesicle length for given $\nu$ and $\lambda$. It is also noticed that this relation gives a more precise estimate of the length for less deflated vesicles. This is because the shape transition point decreases with increasing confinement, as shown in Fig.~\ref{fig_transition}(a). For comparison, the asymptotic theory of Ref.~\cite{barakat_shaqfeh_2018a} showed that $\ell = \ell_c +  O(1 - \lambda/\lambda_c)$ as $\lambda \to \lambda_c$, which is effectively equivalent to Eq.~(\ref{eq:sL}) when $\lambda = \lambda_c$. 

\subsection{Vesicle mobility and extra pressure drop} 
\label{Sec:VP}
\begin{figure}[!htbp]
  \centering
 \includegraphics[scale=0.9]{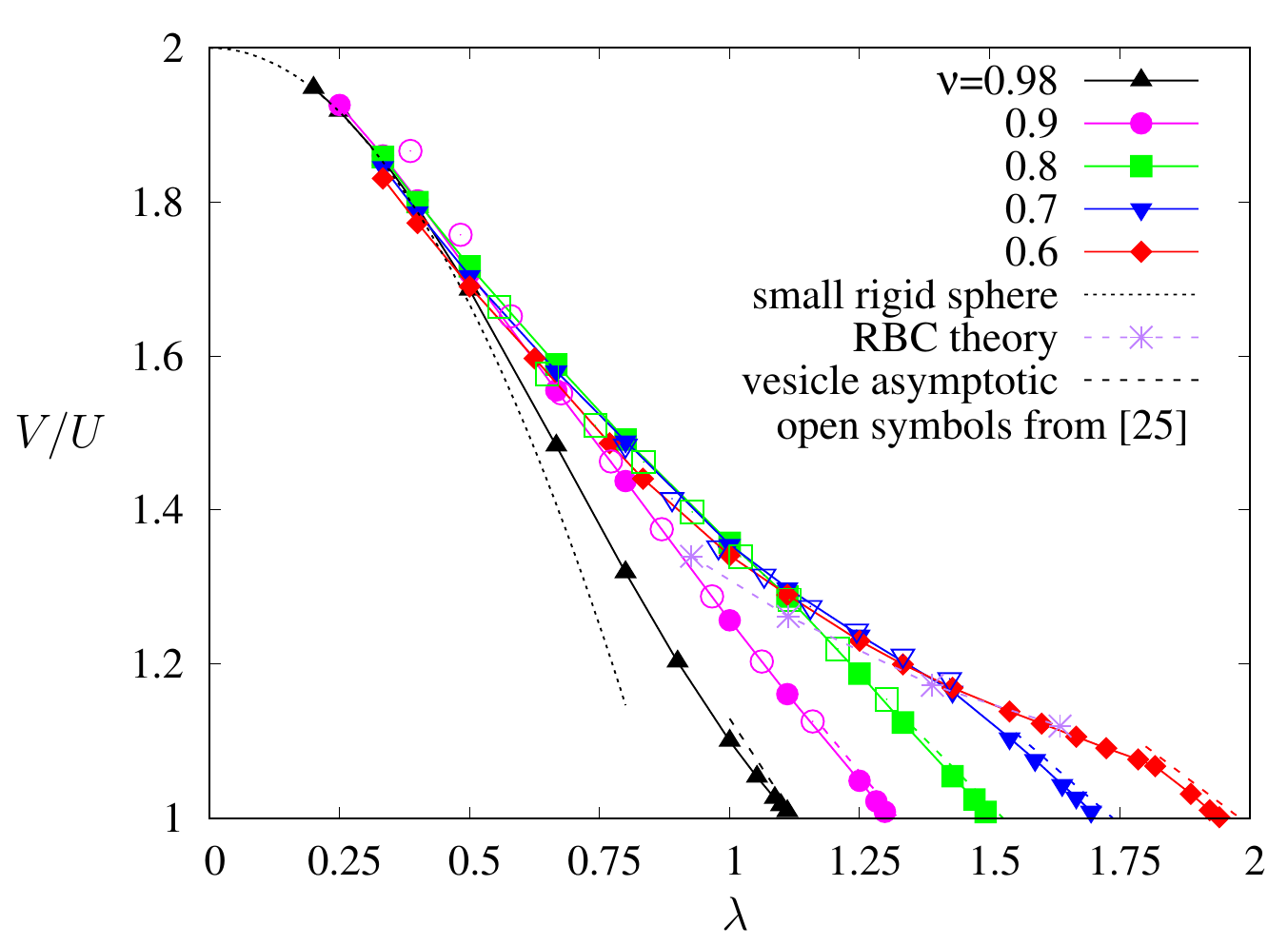}\\
 \caption{Variation of the relative velocity $V/U$ as a function of the confinement $\lambda$ for a wide range of reduced volumes $\nu$ ($\mathrm{Ca_B}=50$). The filled symbols represent the present numerical results with the open symbols of the same shape and color for the 3D numerical results of~\cite{barakat_shaqfeh_2018b} (figure 10, $\nu=0.9$, 0.8, and 0.7). Also shown are the asymptotic prediction for a small rigid sphere moving along the centreline of a tube (\ref{eq:small1}) (dotted line), the asymptotic predictions of~(\ref{eq:asmyp1}) (dashed lines colored differently according to the appropriate reduced volume), and the prediction of a lubrication model for red blood cells~\cite{SecombSkakak1986} (RBC theory).}
	\label{fig_vel}
\end{figure}

\begin{figure}[!htbp]
  \centering
 \includegraphics[scale=0.9]{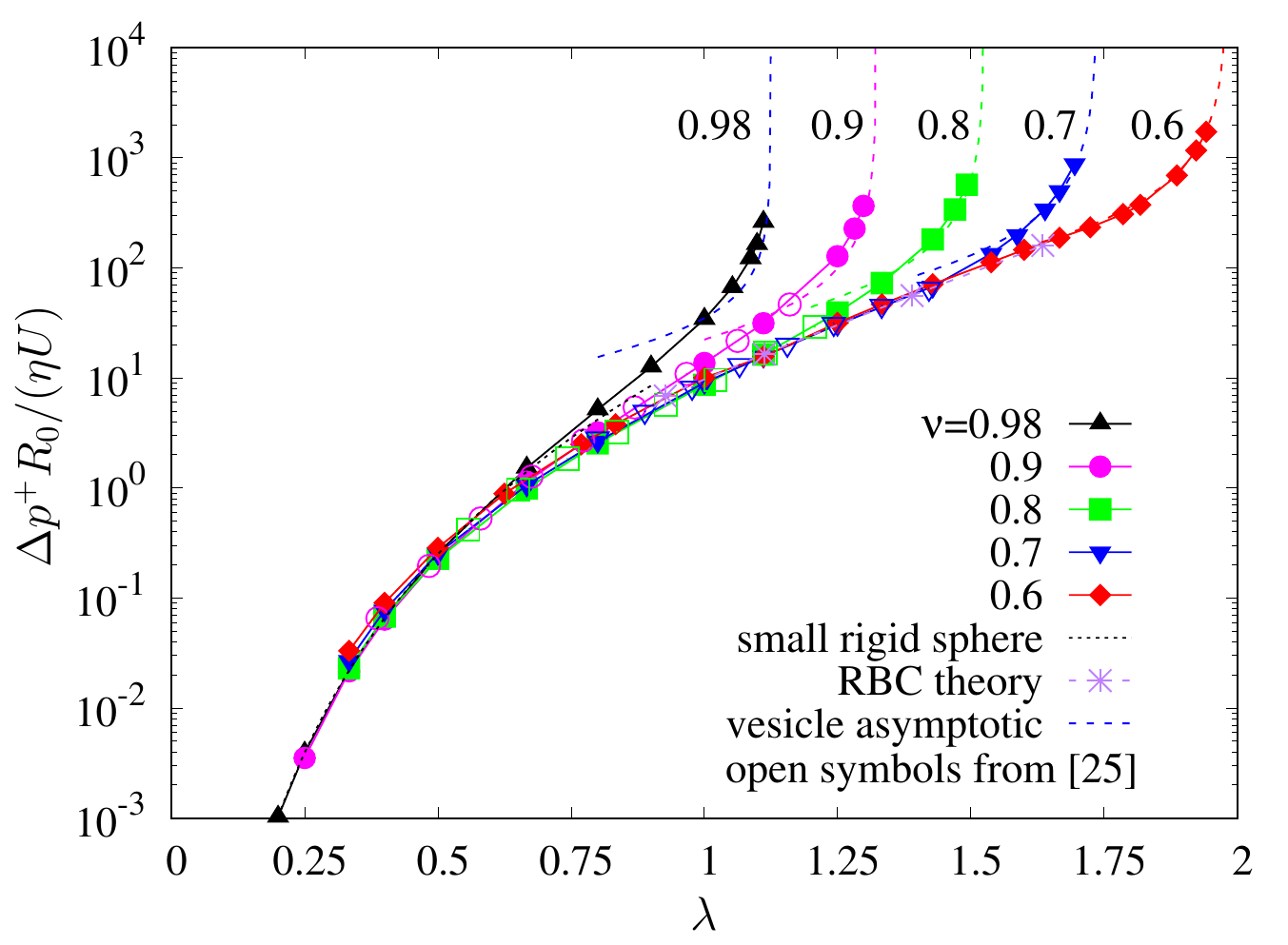}
 \caption{Variation of the dimensionless pressure drop $\Delta p^+ R_0/(\eta U)$ as a function of the confinement $\lambda$ for a wide range of  reduced volumes $\nu$ ($\mathrm{Ca_B}=50$). The filled symbols represent the present numerical results with the open symbols of the same shape and color for the 3D numerical results of~\cite{barakat_shaqfeh_2018b} (figure 11, $\nu=0.9$, 0.8, and 0.7). Also shown are the asymptotic prediction for a small rigid sphere moving along the centreline of a tube (\ref{eq:small2}) (dotted line), the asymptotic predictions of~(\ref{eq:asmyp2}) (dashed lines), and the prediction of a lubrication model for red blood cells~\cite{SecombSkakak1986} (RBC theory).}
 \label{fig_dpa}
\end{figure}

Vesicle mobility, measured in the relative velocity $V/U$, and dimensionless extra pressure drop $\Delta p^+ R_0/(\eta U)$ are the most important hydrodynamical quantities of interest. Especially, the dimensionless extra pressure drop is involved in the determination of the relative apparent viscosity  of a vesicle suspension in tube flow (see~Sec.~\ref{Visco}).  Unlike in an unperturbed Poiseuille flow in which the mean flow velocity $U$ is a linear function of the pressure drop $\Delta p^0$, $V$ and  $\Delta p^+$ are strongly nonlinear coupled due to the vesicle's deformation. Prediction of $V/U$ and $\Delta p^+ R_0/(\eta U)$ has been recently reported in Ref.~\cite{barakat_shaqfeh_2018b} but limited to the reduced volume up to 0.7, presumably due to the difficulty of dealing with the reduced volume of 0.6 in a 3D computation. The present BEM simulations provide a whole range of these quantities in terms of the reduced volume $\nu$ and the confinement $\lambda$, thus extending previous studies of vesicle hydrodynamics in tube flows.  As shown in Fig.~{\ref{fig_vel}} for the relative velocity and Fig.~{\ref{fig_dpa} for the dimensionless extra pressure drop, our simulation results confirm and enhance the numerical results of~\cite{barakat_shaqfeh_2018b}; the results with $\nu=0.6$ (mimicking red blood cells) represent a new addition to Ref.~\cite{barakat_shaqfeh_2018b}. Furthermore, it is clear from these two figures that vesicles with $\nu=0.6$ exhibit a remarkably different behavior as compared to less deflected ones (i.e., $\nu > 0.6$).}

The relative velocity lies between 2 and 1. The former represents an infinitely small vesicle traveling along the tube axis with the centerline maximum flow velocity, and the latter, for a tightly fitting vesicle moving at the mean flow velocity as a "piston" through the tube.  The dimensionless extra pressure drop, however, exhibits large variations -- six orders of magnitude.   Under weak confinement (i.e., small $\lambda$), the simulation results are in excellent agreement with the theoretical predictions for a small spherical droplet flowing along the centreline of a tube~\cite{Hetsroni_JFM_1970,Brenner_1970}, given by

\begin{subequations} \label{eq:small}
 \begin{gather}
 \frac{V}{U}= 2 - \frac{4}{3} \lambda^2 + O(\lambda^3), \label{eq:small1}  \\
 \frac{\Delta p^+ R_0}{\eta U} = 16 \lambda^6   + O(\lambda^{11}) . \label{eq:small2}
 \end{gather}
\end{subequations}
For instance, when $\lambda < 0.5$, the relative errors as compared to the theoretical predictions are less than $1\%$, particularly for nearly spherical vesicles. The case of $\nu=0.6$ is an exception; a decreased mobility and an enhanced pressure drop are clearly visible. These are attributed to the large deformations inherent to the bell-shaped morphology (c.f., Fig.~\ref{fig_shapes2}(d)).  The experimental measurements of $V/U$ reported in Ref.~\cite{{Europhysics_Vitkova_2004}} are not shown herein for the comparison because the measured $V/U$ for vesicles in circular tubes with $\nu=0.924$--$0.999$ are scattered around the curve~(\ref{eq:small1}).

As the confinement increases, the dimensionless groups $V/U$ and  $\Delta p^+ R_0/(\eta U)$ undergo a considerable variation with the reduced volume. Such a high sensitivity to the vesicle's deflation stems from significant changes in vesicle deformations at increasing confinement. Indeed, for a given vesicle, namely a given $\nu$, increasing $\lambda$ results in two combined effects: the vesicle tends to become more elongated, forming a nearly uniform viscous film between the vesicle and the tube wall, as shown in~Fig.~\ref{fig_shapes2} and in~Fig.~\ref{fig_length}, and the gap size becomes smaller, as illustrated in~Fig.~\ref {fig_film}. These two effects enhance the confinement-induced viscous friction on the vesicle surface, thus increasing extra pressure drop across the vesicle, and hindering vesicle mobility. Note that, as derived in the appendix, the shear stress exerted on the membrane is balanced by the tension gradient in the membrane. This is in contrast with a clean drop (i.e., stress-free surface) transported in a pressure-driven flow wherein there appears a plateau value of $V/U$ and  $\Delta p^+ R_0/(\eta U)$ as the confinement increases. Comparisons of the present results with a lubrication model of Ref.~\cite{SecombSkakak1986} for red blood cells are also shown in these two figures; very good agreements are found when $\lambda > 1.4$. For smaller cells (i.e., small $\lambda$), the parallel-flow approximation of the lubrication model produces relatively smaller values of $V/U$ and higher values of $\Delta p^+ R_0/(\eta U)$, which is clearly visible in Fig.~{\ref{fig_vel} but indistinguishable in Fig.~\ref{fig_dpa} due to logarithmic scales used.  When $\lambda \to \lambda_c$, the asymptotic theory of Ref.~\cite{barakat_shaqfeh_2018a} produced, in our notation, the following predictions

\begin{subequations} \label{eq:asmyp}
 \begin{gather}
 \frac{V}{U}= 1 + \frac{4}{3} \left ( \frac{3\lambda_c^2\nu^{-2/3} -2}{4\lambda_c^2\nu^{-2/3}- 3} \right ) \left ( 1 - \frac{\lambda}{\lambda_c}\right ) + O\left [\left ( 1 - \frac{\lambda}{\lambda_c}\right )^2 \right ], \label{eq:asmyp1}  \\
 \frac{\Delta p^+ R_0}{\eta U} = 4\lambda \left ( \lambda_c^2\nu^{-2/3}- 1\right ) \left ( 1 - \frac{\lambda}{\lambda_c}\right )^{-1} +
 \lambda \left ( \frac{4\sqrt{2}\pi}{4 \lambda_c^2\nu^{-2/3} - 3}  \right )\left ( 1 - \frac{\lambda}{\lambda_c}\right )^{-1/2} + O(1) , \label{eq:asmyp2}
 \end{gather}
\end{subequations}

The limiting behaviors are well captured in the simulations. We note that the numerical results of $V/U$ are slightly less than those predicted by the theory as $\lambda \to \lambda_c$, and the discrepancy becomes visible at low reduced volumes, particularly at $\nu=0.6$. Two potential reasons might explain the discrepancy shown in Fig.~\ref{fig_vel} between the asymptotics and numerics. Firstly, it may be due to the assumed fore-aft symmetry of a sphero-cylindrical shape when calculating the critical confinement $\lambda_c$. The frontal endcap is relatively smaller than the rear endcap due to an outward bulge there, as shown in Fig.~\ref{fig_app}, resulting in a slightly larger theoretical $\lambda_c$. The outwardly bulging endcap becomes noticeable with increasing vesicle's deflation, leading to a more marked discrepancy at low reduced volumes. The second reason may be a simple numerical issue (despite we have checked all simulations with higher resolution, the numerical results remain unchanged) since the comparison between numerics and asymptotics presented in Ref.~\cite{barakat_shaqfeh_2018b} shows no such discrepancy. Further work is still needed to clarify this issue. 

Another noticeable feature is a sudden change of slope in $V/U$; it becomes visible for $\nu=0.8$ and $0.7$, and that feature is quite noticeable for $\nu=0.6$. This change in slope was also pointed out for $\nu=0.7$ in the 3D BEM simulations of~\cite{barakat_shaqfeh_2018b}. We found that the points where the curves of $V/U$ vs. $\lambda$ display a remarkable change in slope correspond exactly to the parachute-bullet transition points in the ($\nu$, $\lambda$) space, as discussed above. This result is not surprising given that vesicle mobility is dictated by the vesicle shape and its interplay with the surrounding fluid. Indeed, the competition between geometric constraints and confinement-induced viscous friction determines the speed at which a vesicle is transported in a pressure-driven flow. 

\begin{figure}[!htbp]
  \centering
 \includegraphics[scale=0.8]{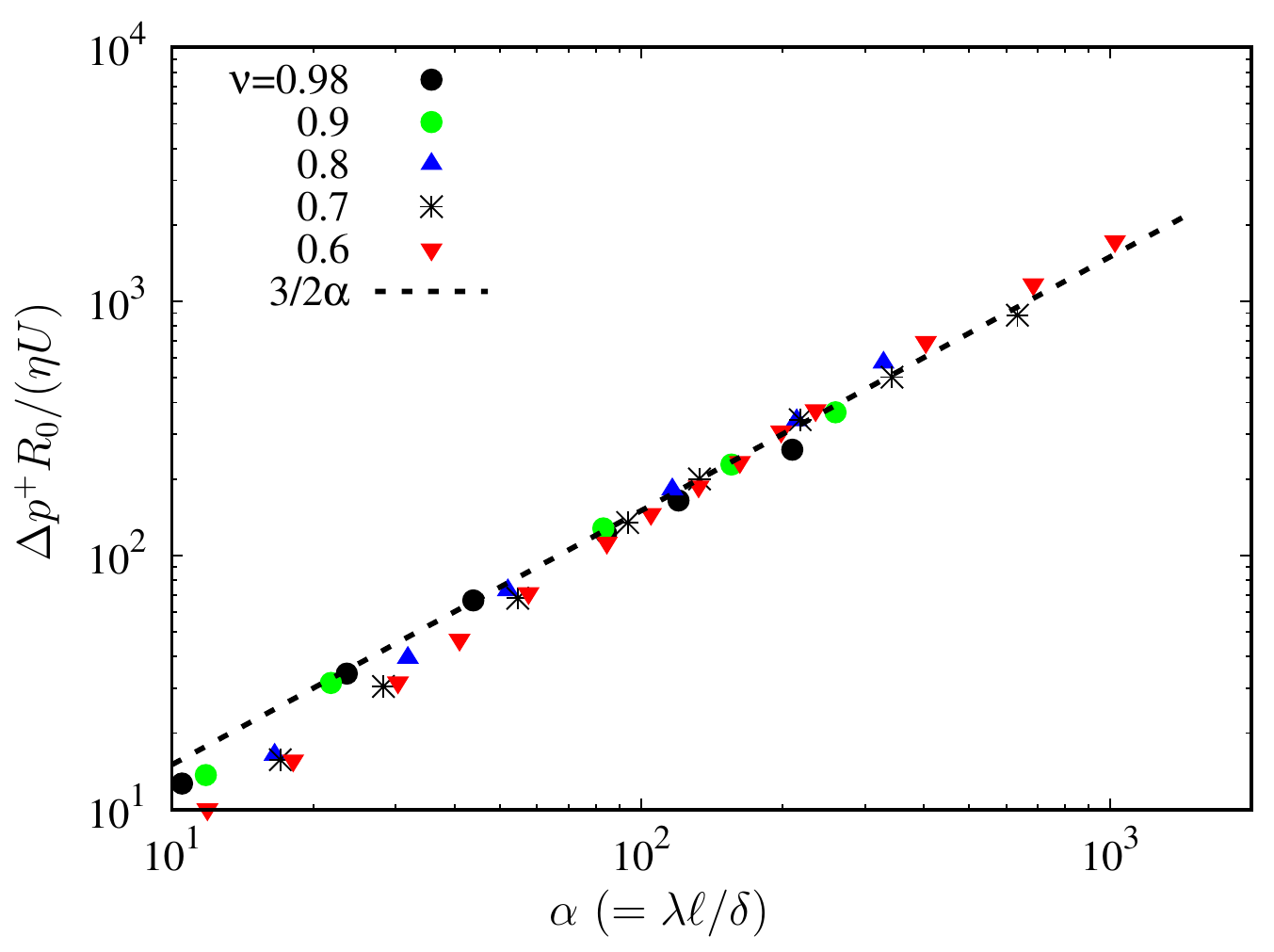}
 \caption{Dimensionless pressure drop $\Delta p^+ R_0/(\eta U)$ plotted as a function of $\alpha~(=\lambda \ell/\delta)$ for highly confined vesicles (i.e., $\lambda \to \lambda_c$). The dashed line shows the scaling $3\alpha/2$. }
	\label{fig_dpa_film}
\end{figure}

\begin{figure}[!htbp]
  \centering
 \includegraphics[scale=0.8]{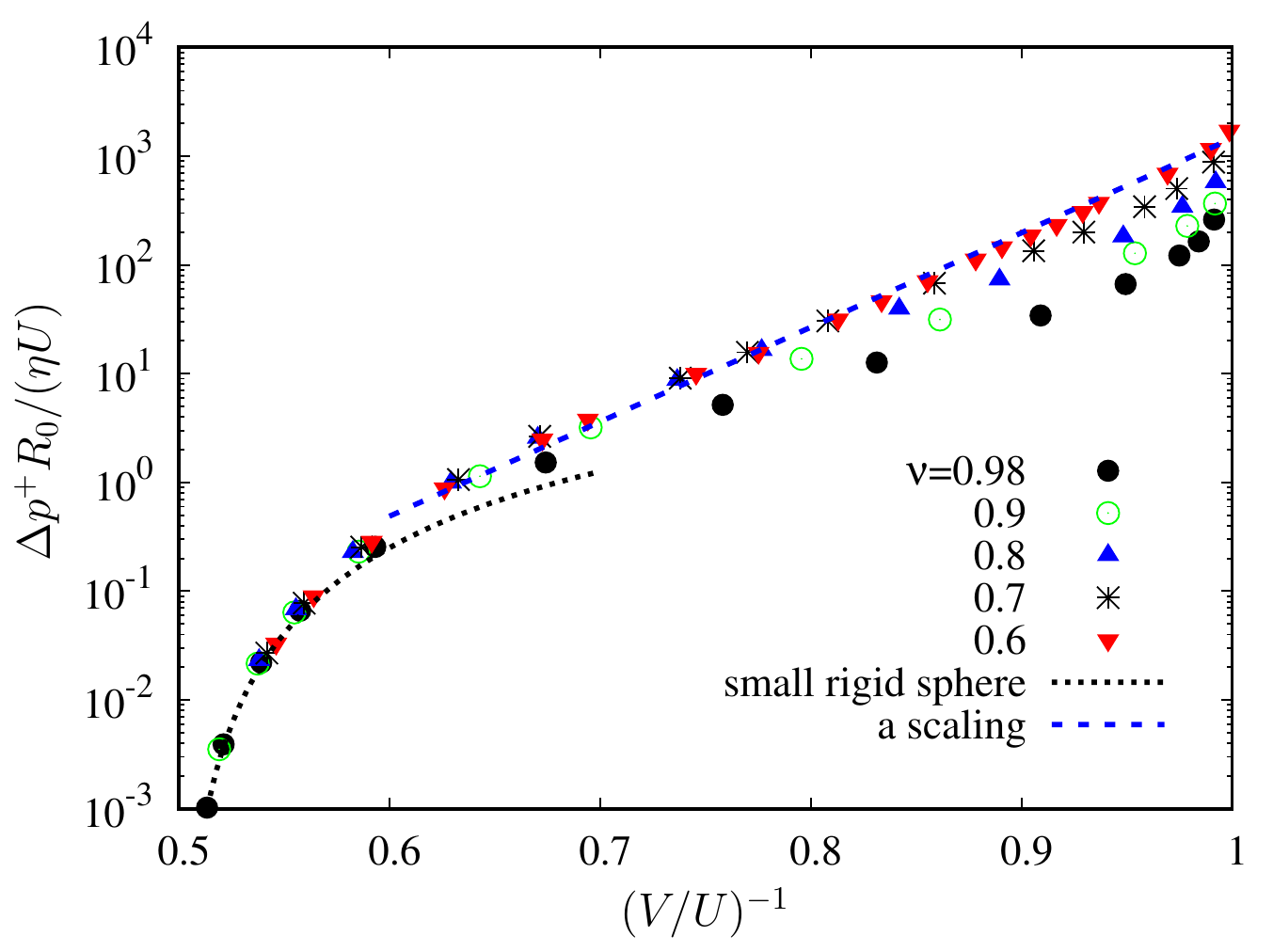}
 \caption{Dimensionless pressure drop $\Delta p^+ R_0/(\eta U)$ versus the reciprocal of the relative velocity $V/U$.  The dotted curve shows the analytical prediction for a small rigid sphere (\ref{eq:dpa_v}). The dashed line is the best fitting (\ref{eq:dpa_v_fit}) to $\nu=0.6$ in the range of $0.6 < (V/U)^{-1} < 0.995$.}
	\label{fig_dpa_v}
\end{figure}

When $\lambda \to \lambda_c$, the asymptotic theory of Ref.~\cite{barakat_shaqfeh_2018a} shows a scaling of the dimensionless extra pressure drop $\Delta p^+ R_0/(\eta U) \sim O(\lambda\ell/\delta) $; the dimensionless extra pressure drop is proportional to the reduced vesicle length but inversely proportional the clearance size. The compilation of the present simulation results allows a precise correlation  $\Delta p^+ R_0/(\eta U) \simeq 3/2 (\lambda \ell /\delta)$, as illustrated in Fig.~\ref{fig_dpa_film}. This is a significant improvement, given that a wide range of  reduced volumes is involved. More importantly, this correlation holds implications that may help devise and interpret tube-flow experiments. Specifically, based on the vesicle length and its translational velocity which are the most easily accessible quantities in experiments, the simulated results presented in Figs.~\ref{fig_length} and \ref{fig_vel}, together with scaling laws obtained from this study allow an estimate of the reduced volume $\nu$ (hence $\lambda_c$) and the film thickness, from which the dimensionless extra pressure drop $\Delta p^+ R_0/(\eta U)$ can be inferred. Directly measuring these parameters is no simple task. It often requires advanced imaging methods and instrumentation, with an added difficulty arising from the fact that the extra pressure drop is highly sensitive to the reduced volume.  The estimated extra pressure drop should be contrasted with Fig.~\ref{fig_dpa} for consistency. An iterative process may be required to obtain a consistent result. Finally, an estimate of the maximum tension in the membrane can also be obtained using the scaling laws shown in Fig.~\ref{fig_film_ca}.

Before closing this subsection, we highlight the strong coupling behavior in the dimensionless groups $\Delta p^+ R_0/(\eta U)$ and $V/U$. Even for a vanishing small rigid sphere, Eqs.~(\ref{eq:small1}) and~(\ref{eq:small2}) give a highly nonlinear relationship
\begin{equation} \label{eq:dpa_v}
 \frac{\Delta p^+ R_0}{\eta U} = \frac{27}{4} \left ( 2 -  \frac{V}{U}(\lambda) \right )^3   + O(\lambda^{3}) .
\end{equation}
The theoretical prediction is shown in Fig.~\ref{fig_dpa_v}, where the dimensionless pressure drop $\Delta p^+ R_0/(\eta U)$ is plotted against the reciprocal of the relative velocity $V/U$. The reason for using $(V/U)^{-1}$, instead of $V/U$, is quite simple and it is to illustrate how the coupling behaves as the vesicle size -- equivalently the confinement for a given tube diameter -- increases. While both the relative velocity and dimensionless extra pressure drop are notably sensitive to the reduced volume as the confinement increases, Fig.~\ref{fig_dpa_v} makes it clear that the sensitivity becomes relatively weaker as compared to Fig.~\ref{fig_dpa}. Nevertheless, finding a correlation between the dimensionless groups $\Delta p^+ R_0/(\eta U)$ and $V/U$, under high confinement, is by no means trivial because $\Delta p^+ R_0/(\eta U)$ diverges like $(1-\lambda/\lambda_c)^{-1}$. So here we put forward only a fitting to $\nu=0.6$, given by
\begin{equation} \label{eq:dpa_v_fit}
 \frac{\Delta p^+ R_0}{\eta U} \simeq 3 \times 10^{-6} \exp \left [ 20 \left (\frac{V}{U} \right )^{-1} \right ] .
\end{equation}
We conclude that the dimensionless groups $\Delta p^+ R_0/(\eta U)$ and $V/U$ are strongly nonlinear coupled in tube flows. However, it is worth emphasizing that both the pressure drop and velocity ratio are measured quantities; neither are controlled parameters.

\subsection{Implications for the rheology of dilute red blood cell suspensions} \label{Visco}

The hematocrit measures the volume of red blood cells (RBCs) compared to the total blood volume (red blood cells and plasma). Its normal value in humans is approximately 45\% but can be largely less in small vessels. Consider now the hematocrit, denoted by $H_T$, in a capillary of length $L$, and assuming RBCs flow in single file through the capillary with a characteristic length $l_v$ between two neighbors (i.e., the distance between their centers of mass), then the ratio $L/l_v$ is the number of RBCs inside the capillary, the total volume of the RBCs is $(L/l_v)\Omega$ ($\Omega$ the volume of a single RBC) and the hematocrit $H_T$ equals to  $(\Omega L/l_v)/(\pi R^2 L)$, which gives a mean distance between RBCs 
\begin{equation}
l_v= \frac{\Omega}{\pi R^2 H_T}.
\end{equation}

Poiseuille's law defines an apparent viscosity in terms of the overall pressure drop across the capillary tube
\begin{equation}
\eta_\mathrm{app} = \left (\Delta p^0 + \Delta p^+\right ) \frac{\pi R^4}{8 L Q} = \eta + \frac{R^2}{8 U L} \Delta p^+ .
\end{equation}
We then obtain, by setting $L=l_v$,
\begin{equation}
\eta_\mathrm{app} =  \eta \left [1 +  \frac{3}{32} \left (\frac{R}{R_0}\right )^4 \left (\frac{R_0}{\eta U} \Delta p^+\right ) H_T \right ] = \eta (1 + K_T H_T) ,
\end{equation}
where the dimensionless parameter $K_T$ is called apparent intrinsic viscosity. This single-file flow model allows recovering an apparent viscosity which depends linearly on the local hematocrit $H_T$, as in a lubrication model for red blood cells~\cite{SecombSkakak1986}. In capillaries with diameters up to about 8 $\mu$m, the single-file flow model is appropriate as RBCs frequently flow in single file and interactions between cells may be negligible.

Finally, the apparent  intrinsic viscosity is given by
\begin{equation}
K_T= 9.53\times10^{-5}  d^{4} \left (\frac{R_0}{\eta U} \Delta p^+\right ) , \qquad \mbox{for} \quad d > d_c,
\end{equation}
where $d$ denotes the diameter of tubes, $d_c$ the critical diameter ($\simeq 2.8~\mu$m). The dimensionless extra pressure drop $\Delta p^+ R_0/(\eta U)$ for $\nu=0.6$ is plotted in Fig.~\ref{fig_dpa} as a function of the confinement $\lambda=R_0/(d/2)$ with $R_0 \simeq 2.8~\mu$m. Relative apparent viscosity which is the ratio of apparent viscosity to suspending medium viscosity can be written in terms of tube diameter $d$ in $\mu$m, dimensionless extra pressure drop, and hematocrit $H_T$ as 
\begin{equation}
\eta_\mathrm{rel} = \frac{\eta_\mathrm{app}}{ \eta} \simeq 1 + 9.53\times 10^{-5}  d^{4} \left (\frac{R_0}{\eta U} \Delta p^+\right ) H_T, \qquad d~~\mbox{in}~\mu\mbox{m}.
\end{equation}

\begin{figure}[!htbp]
  \centering
 \includegraphics[scale=0.8]{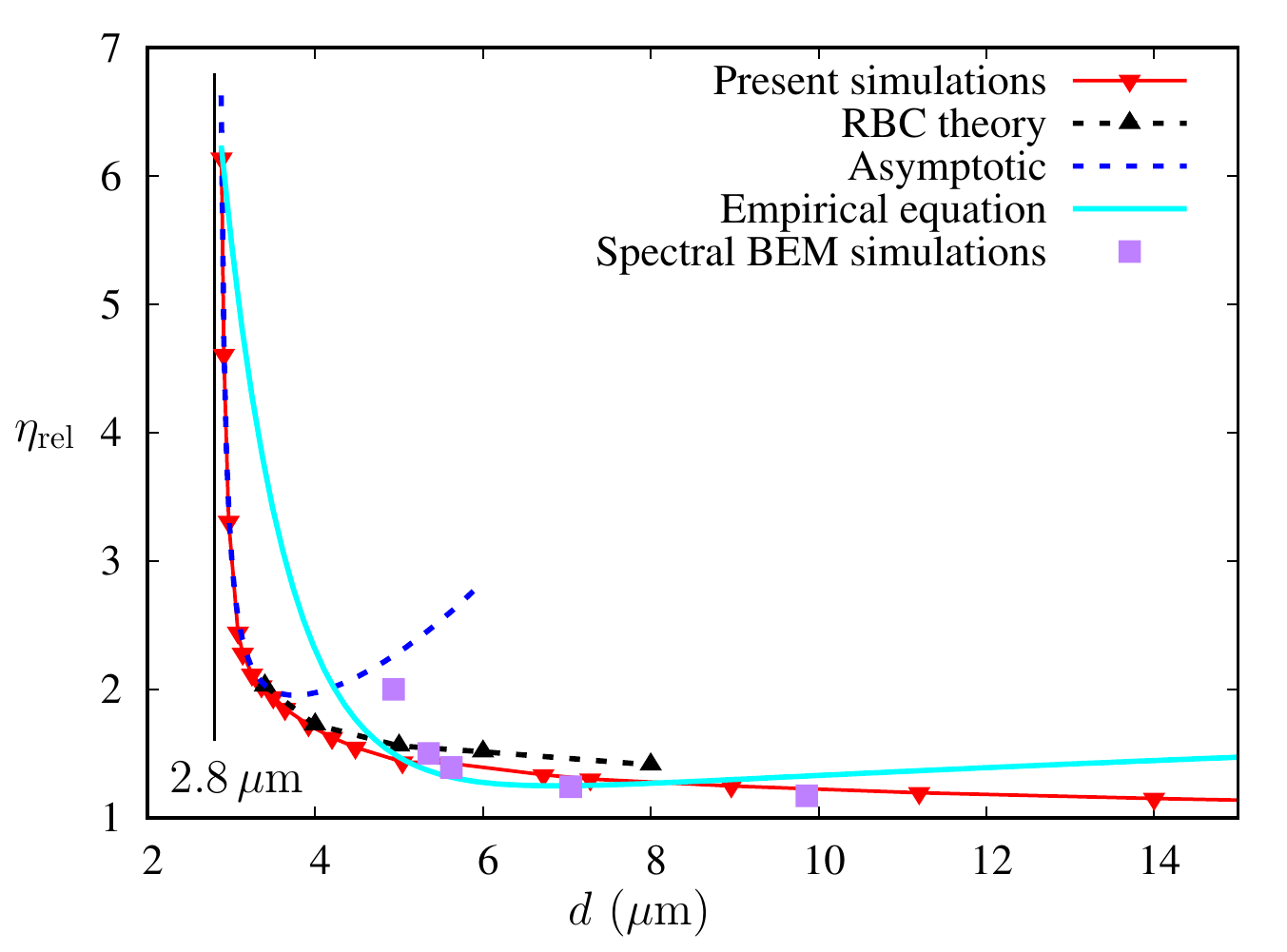}\\
 \caption{Variation of relative apparent blood viscosity $\eta_\mathrm{rel}$ with tube diameter $d$ in $\mu$m for a hematocrit $H_T$ of 0.45. Curve --\textcolor{red}{{$\blacktriangledown$}}-- represents simulation results based on single-file fluid model, and curve --\textcolor{black}{{$\blacktriangle$}}-- represents lubrication model of RBCs~\cite{SecombSkakak1986}. Dashed line in blue shows asymptotic theory for dimensionless extra pressure drop~\cite{barakat_shaqfeh_2018a} (Eq.~(\ref{eq:asmyp2})).  Solid curve in cyan represents a fitting empirical equation to  {\it in-vitro} experimental data~\cite{Pries_1992}. Also shown is the spectral 3D BEM simulations of Ref.~\cite{Zhao_JCP_2010} (for $H_T=0.30$). The vertical line (black) indicates a lower limit ($d_c\simeq 2.8$ $\mu$m) to the diameter of tubes beyond which normal red blood cells cannot pass through without rapture.}
	\label{fig_Kt}
\end{figure}

As an example, simulated relative apparent viscosity with a hematocrit of 0.45 as a function of tube diameter is shown in Fig.~\ref{fig_Kt}. There is a small decrease in $\eta_\mathrm{rel}$ as the tube diameter increases when $d\ge 5~\mu$m (i.e., $\lambda \le 1.11$). The value of $\eta_\mathrm{rel}$ for $d = 5.6~\mu$m ($\lambda=1$) is slightly less than $2\%$ of the value for $d = 5~\mu$m. At smaller diameters, say $d<4~\mu$m, relative apparent viscosity rises rapidly and becomes substantially higher as the tube diameter approaches the critical diameter $d_c$. The dramatical rise is attributed to a significantly large resistance to flow as reflected from the behavior of extra pressure drop at nearly maximum confinement. This feature is remarkably captured in the asymptotic theory of Ref.~\cite{barakat_shaqfeh_2018a} for the tube diameters less than $3.5~\mu$m.  The numerical results are compared with those obtained from a lubrication model of red blood cells at high shear rates ($U/d > 50~\mathrm{s}^{-1}$)~\cite{SecombSkakak1986}. It is shown that apparent viscosity is almost independent of flow rate in this regime but increases with decreasing flow rate at lower shear rates~\cite{SecombSkakak1986,Pries_1992}. At a bending capillary number $\mathrm{Ca_B}=50$, our BEM simulations always lie in a high-flow-rate regime as an estimate of $U/d =  \kappa \mathrm{Ca_B} \lambda  /(2 \eta R_0^3) > 50~\mathrm{s}^{-1}$ even at a very weak confinement $\lambda=0.3$. While the model of Ref.~\cite{SecombSkakak1986} includes a shear elasticity of the RBC membrane but neglects the bending elasticity, our numerical predictions of $\eta_\mathrm{rel}$ are in excellent agreement with the lubrication theory when $d \le 4~\mu$m. Indeed, under high confinement, bending resistance has a negligible contribution to the hydrodynamic force balance; the isotropic tension in the membrane (see Appendix) resists the flow in the lubrication layer. When $d>5~\mu$m, the parallel-flow approximation of Ref.~\cite{SecombSkakak1986} yields relatively higher values of $\eta_\mathrm{rel}$ as compared to our BEM simulation results.  As shown, the present simulation results are also in very good agreement with the spectral 3D BEM simulations reported in Ref.~\cite{Zhao_JCP_2010} except at a very small tube diameter. 

Based on a compilation of {\it in-vitro} experimental data, an empirical equation describing the dependence of relative apparent viscosity on tube diameter has been put forward in Ref.~\cite{Pries_1992} and is also plotted in Fig.~\ref{fig_Kt}. Given the paucity of experimental measurements in the range of smaller tube diameters, we may say that the predicted relative apparent viscosities are in reasonable agreement with experimental data for tube diameters ranging between 2.9 and 14~$\mu$m. Nevertheless, it should be mentioned that while the present single-file vesicle model provides some insight into how apparent blood viscosity behaves for tube diameters in the range of $\sim 2.8$--14~$\mu$m, the model due to its axial symmetry nature is not able to make reliable predictions for tube diameters beyond that range as the confinement ($\lambda < 0.4$) becomes too weak for vesicles to preserve axisymmetry. Also, the limitation of a single-file flow model (i.e., $d < 8~\mu$m) makes the model unreliable for the prediction of relative apparent viscosity for larger tube diameters; the simulated results presented in Fig.~\ref{fig_Kt} for tube diameters larger than 8~$\mu$m are for illustrative purposes only.

\section{Summary and concluding remarks} \label{Conclusions}

We have presented a numerical investigation of the motion and deformation of a vesicle freely suspended inside a circular tube in a pressure-driven flow.
The numerical simulations of this fluid-structure interaction problem have been carried out by using a previously reported axisymmetric boundary element method. The results were presented for the reduced volumes $\nu$ ranging from 0.6 (i.e., red blood cell-mimicking vesicles) to 0.98 (i.e., nearly spherical vesicles) at different degrees of confinement, up to near its critical value $\lambda_c$. The critical confinement of a vesicle in cylindrical tube flow, as well as its critical length $\ell_c$, was calculated on the basis of its physical constraints of fixed volume and surface area while assuming the fore-aft symmetry of a sphero-cylindrical shape. 

The numerical results allowed us to build a phase diagram of vesicle shapes in good agreement with the most comprehensive experimental data reported in the literature~\cite{Coupier_PhysRevLett108_2012} . Carefully controlled simulations let us establish a linear shape transition line separating the two commonly observed shapes, namely parachute-like and bullet-like shape regions in the ($\lambda$, $\nu$) space.  We found that the shape transition is accompanied by a change in the  behavior of the mobility of vesicles, especially for low-reduced-volume vesicles (i.e., $\nu \le 0.7$).  The present work focused on highly confined vesicles, which required high-resolution simulations to accurately compute vesicle shapes, membrane traction, and wall resistance. These simulations enabled us to examine the limiting behavior of several quantities of interest when $\lambda \to \lambda_c$, particularly the vesicle mobility $V/U$ and the dimensionless extra pressure drop $\Delta p^+ R_0/(\eta U)$ due to the presence of the vesicle in the tube. Our numerical results lend support to a recently reported asymptotic theory~\cite{barakat_shaqfeh_2018a}. 

Aiming to help interpret the numerical results when the confinement approaches its critical value, we have also presented a lubrication theory combining two approaches described in the literature. While the balance between viscous, bending and tension forces controls the vesicle motion and deformation, we showed that bending elasticity plays a minor role in the force balance in the lubrication layer. It is the tension gradient along the membrane that resists the confinement-induced viscous friction, thus controlling the size of the gap between the tube wall and the vesicle surface.  We found that the maximum membrane tension in RBC-like vesicles at the proximity of the critical confinement is on order of $\sim$ 1~mN/m -- which approaches the rupture tension. 
We should point out that several previous studies (e.g., Refs.~\cite{SecombSkakak1986,Bruinsma1996,barakat_shaqfeh_2018a,barakat_shaqfeh_2018b}) have concluded that bending resistance has a secondary effect on the film thickness at high flow rates.

Based on a single-file flow model, an attempt has been made to predict the rheology of dilute red blood cell suspensions. Simulated relative apparent viscosity of a vesicle suspension in small capillary tubes yielded a consistent and complementary result as compared with experimental data and highlighted the role of confinement in the rapid rise in the relative viscosity of red blood cells when passing through small vessels. There is, however, a severe limitation to such a  model since it is relevant only for capillary diameters up to about 8~$\mu$m.

Being able to deal with a wider range of reduced volumes, the present BEM simulations extend previously published studies on the hydrodynamics of a vesicle in tube flows and elucidate the intricate interplay between the deformation of the vesicle and its mobility. The numerical results, together with various correlations gained, may help to predict some parameters that are difficult to measure in tube-flow experiments. For instance, the extra pressure drop, which is the most relevant parameter for the rheology of a dilute suspension of vesicles, can be inferred from the easily accessible parameters, such as the vesicle velocity and its length.  It is our hope that the results presented in this paper could serve as a benchmark for future studies and help devise and interpret tube-flow experiments.

We acknowledge that despite the conclusions from the present study, there is one major simplifying assumption that will need to be revisited in future work. We limited our model to axisymmetric shapes, while it may be true in highly confined capillary flow, vesicles exhibit non-axisymmetric shapes as well~\cite{Kaoui_PRL2009,Farutin_PRE_2011,Coupier_PhysRevLett108_2012,Quint2017}.  In their 3D BEM simulations, Barakat and Shaqfeh~\cite{barakat_shaqfeh_2018b} highlight the emergence of asymmetric and time-dependent vesicle shapes in tube flow, especially at low confinement and low reduced volumes. The 3D simulations of Zhao et al.~\cite{Zhao_JCP_2010} of RBCs in tube flow also show that RBCs break axisymmetry. Future studies will involve performing fully three-dimensional computer simulations of confined vesicles. These simulations will provide a more complete picture of vesicle shape diagrams and unveil its complex dynamics in confined channel flow. 

\begin{acknowledgments}
We acknowledge financial supports from Labex MEC (grant no. ANR-11-LABX-0092), from A*MIDEX (grant no. ANR-11-IDEX-0001-02), from ANR (grant no. ANR-18-CE06-0008-03), from the LabEx Tec21 (ANR-11-LABX-0030), from the PolyNat Carnot Institute (ANR-11-CARN-007-01) and from CNES. J.M. Lyu was sponsored by the China Scholarship Council (CSC). Centre de Calcul Intensif d’Aix-Marseille is acknowledged for granting access to its high performance computing resources. We thank the anonymous reviewers for their many insightful comments and suggestions, which helped to improve this paper.
\end{acknowledgments}

\appendix*
\section{Lubrication theory}

We combine previously reported results in the literature~\cite{Lighthill_1968,SecombSkakak1986,Bruinsma1996,barakat_shaqfeh_2018a,barakat_shaqfeh_2018b} to present an axisymmetric form of lubrication theory for a vesicle in tube flow. Our aim is not to numerically resolve the complete system of governing equations of the lubrication theory, as it was generally conducted by those authors. Instead, we make use of some asymptotic scaling laws established via lubrication theory analysis to help interpret the present simulation results when confinement is close to its maximum value (i.e., $\lambda \to \lambda_c$). 
The following paragraphs are an attempt to combine two approaches, one is based on parallel-flow approximation ~\cite{SecombSkakak1986,Bruinsma1996,barakat_shaqfeh_2018b}, the other is the small-gap theory in the singular limit $\lambda \to \lambda_c$~\cite{barakat_shaqfeh_2018a}. Starting from the well-established theories developed in~\cite{Lighthill_1968,Hochmuth_1970}, we show that the two approaches lead to the same asymptotic behavior of the film thickness and vesicle velocity in the limit $\lambda \to \lambda_c$, and the film thickness is further controlled dynamically by the membrane tension.

\begin{figure}[!ht]
  \centering {\includegraphics[scale=1.2]{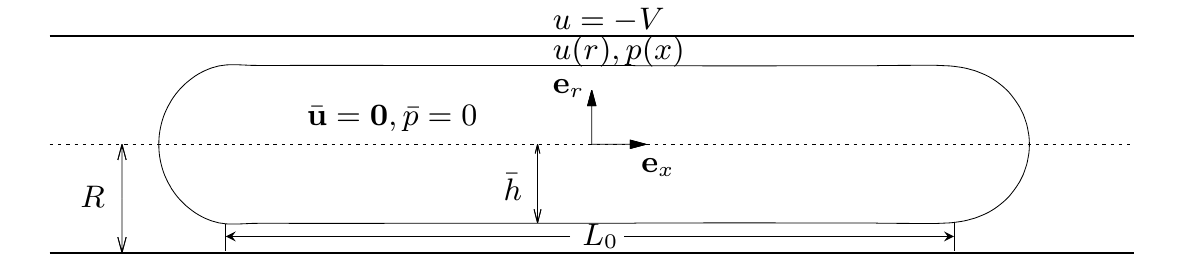}}
 \caption{Schematic of lubrication theory analysis for a steady, sphero-cylindrical  vesicle moving along the axis of a circular tube of radius $R$. In a coordinate frame moving with the vesicle, there can be no flow and pressure gradient inside the vesicle. The vesicle has a cylindrical main body with a length $L_0$, assumed large compared to $R$. Plotted vesicle profile corresponds to a vesicle shape for $\nu=0.6$, 
 $\lambda=1.92$ and $\mathrm{Ca}_\mathrm{B}=50$ (shown in Fig.~\ref{fig_shapes2}(d)); the tube wall is placed at more than $20\%$ away from its actual place in order to amplify the width of the gap between the vesicle and the tube wall.}
 \label{fig_app}
\end{figure}

A schematic description of lubrication theory analysis is shown in Fig.~\ref{fig_app}. The shape of the membrane is described by $r=\bar{h}(x)$, and the gap separating the vesicle membrane and tube wall has a typical film thickness $h$, which is defined at the vesicle's midplane, namely $h \equiv R-\bar{h}(0)$. At this stage, we assume that the thickness is small compared to the vesicle length $L$ (but not necessarily small relative to the tube radius), as it is usually the case with high confinements (e.g., as illustrated in Fig.~\ref{fig_shapes2}). In the reference frame moving with a steady vesicle centered along the axis of the tube, the axial symmetry of the problem and the incompressibility of the vesicle membrane lead to 
\begin{itemize}
\item the fluid inside the vesicle  is stationary with no pressure gradient (for simplicity we set $\bar{p}=0$) and behaves like a rigid-body ($\bar{\mathbf{u}}=\mathbf{0}$);
\item the viscous normal stress on the membrane vanishes 
%(since  $\mathbf{n} \cdot  (\bm{\sigma} - p\mathbf{I}) \cdot \mathbf{n} = \nabla_\mathrm{S} \cdot \mathbf{u} = 0$), 
so that the net normal traction on the membrane is the hydrostatic pressure difference between the internal and external flows $\bar{p} -p $, which is equivalent to be $-p$;
\item  the net shear traction is the viscous shear stress $\tau$ due to the external fluid. 

\end{itemize}
 Then, in the lubrication approximation, the pressure $p$ in the lubrication layer depends only on the axial position $x$, and the axial velocity field $u$ is governed by the axial momentum equation and an equation of continuity

\begin{subequations} \label{eq:lub}
 \begin{gather}
 \frac{\eta}{r}\frac{\partial }{\partial r} \left (r \frac{\partial u}{\partial r} \right)=\frac{\mathrm{d}p}{\mathrm{d}x}, \label{eq:lub1} \\
 \int_{\bar{h}(x)}^{R} 2 \pi u r \mathrm{d}r = \pi R^2 (U - V) \equiv -2 \pi R q , \label{eq:lub2}
 \end{gather}
\end{subequations}
subject to boundary conditions
\begin{subequations} \label{eq:lubb}
  \begin{gather}
u = -V \qquad  \mbox{at} \quad r=R, \label{eq:lubb1} \\
u = 0  \qquad  \mbox{at} \quad r=\bar{h}(x), \label{eq:lubb2}  
\end{gather}
\end{subequations}
where $q$~($=\frac{1}{2} R (V-U)$) represents a rate per unit circumference  of leakback of fluid past the vesicle.

 Equations~(\ref{eq:lub}) with boundary conditions~(\ref{eq:lubb}) yield the Reynolds lubrication equation~\cite{Hochmuth_1970} for the pressure gradient in terms of the azimuthal radius of curvature $r(x)= \bar{h}(x)$:
\begin{equation}
\frac{\mathrm{d}p}{\mathrm{d}x} = - \frac{8\eta U}{R^2 \left ( 1 - (r/R)^2\right )}\left[ \left ( \frac{1 - (r/R)^2}{2\log(r/R)}\right) \frac{V}{U}  -  \frac{2 q}{R U}\right] 
 \left[ 1 + (r/R)^2 + \frac{1 - (r/R)^2}{\log(r/R)}\right]^{-1} .
\label{eq:lubp}
\end{equation}
The shear stress exerted on the membrane due to the ambient flow is then given by
\begin{equation}
\tau(r) \equiv \eta \frac{\partial u}{\partial r} = -\frac{1}{4} \frac{\mathrm{d}p}{\mathrm{d}x}  \left[ 2r + \frac{R^2 - r^2}{r \log(r/R)}  \right] +  \frac{\eta V}{r \log(r/R)}  .
\label{eq:shear}
\end{equation}

The normal and shear stress boundary conditions~(\ref{eq:dynam}), together with (\ref{E:fm}), respectively, can now be approximated by
\begin{equation}
\label{lub:norm}
- p = - 2\kappa \left [ \frac{1}{g_s} \frac{\mathrm{d}  }{\mathrm{d}x} \left (  \frac{r^2}{g_s} \frac{\mathrm{d} H }{\mathrm{d}x}\right)  + 2 H(H^2 - K) \right ]  + 2\gamma H, 
\end{equation}
\begin{equation}
 \tau = - \frac{r}{g_s} \frac{\mathrm{d} \gamma }{\mathrm{d}x},
 \label{eq:tau}
\end{equation}
 where $g_s = r \sqrt{1+(\mathrm{d}r/\mathrm{d}x)^2}$ is the surface metric. The mean  and Gaussian curvatures can be written in terms of $r(x)$ and its derivatives:
\begin{subequations}\label{eq:cur}
 \begin{gather}
 H =  \frac{1}{2} (c_1 + c_2), \qquad K = c_1 c_2, \\
 \mbox{with} \qquad c_1 = \frac{1}{g_s}, \quad c_2 = - \frac{r^3}{g_s^3} \frac{\mathrm{d}^2 r }{\mathrm{d}x^2}.
\end{gather}
\end{subequations}
 
 Equations~(\ref{eq:lubp})--(\ref{eq:tau}), together with the usual symmetry conditions at the front nose of the vesicle $x=L/2$ and at the rear tail of the vesicle $x=-L/2$, are solved numerically in~\cite{SecombSkakak1986,barakat_shaqfeh_2018b}. It is shown that the above axisymmetric lubrication equations yield effectively good approximations to the Stokes flow of a vesicle inside a circular tube if the membrane slope $|\mathrm{d}r/\mathrm{d}x|$ is sufficiently small. 

To gain insight into the situation of narrow gaps, we now make an additional assumption that the thickness of the lubricating film between the membrane and inner tube wall is small relative to the tube radius. In this case, the leakback is also small compared with the total flow. Introducing a small parameter 
\begin{equation}
 \epsilon \equiv   \frac{2q}{UR} = \frac{V}{U} - 1 \ll 1,
 \label{eq:epsilon}
\end{equation}
and a rescaled film thickness $h^*$ such that
\begin{equation}
 r =R(1-\epsilon h^*),
 \label{eq:hstar}
\end{equation}
we obtain approximate solutions for the pressure gradient and shear stress~\cite{Lighthill_1968,Hochmuth_1970} and their simplified forms in a lubrication layer of uniform thickness ($h$) with a pure shear flow:
\begin{subequations}\label{eq:approx}
 \begin{gather}
 \frac{\mathrm{d}p}{\mathrm{d}x}  = - \frac{6\eta U}{(R\epsilon h^*)^2} \left [ \frac{V}{U} -  \frac{1}{h^*} + O(\epsilon h^*) \right ] = 
 - \frac{6\eta}{h^2} \left [ V -  \frac{2q}{h} + O(\epsilon) \right ], \label{eq:approx1} \\
\tau =  \frac{\eta U}{R\epsilon h^*} \left [ 2 \frac{V}{U} - \frac{3}{h^*} + O(\epsilon h^*) \right ] =  
 \frac{2\eta}{h} \left [ V -  \frac{3q}{h} + O(\epsilon) \right ] \label{eq:approx2} .
%\frac{2\eta V}{\hbar}  - \frac{6\eta q}{\hbar^2} + O(\epsilon) \label{eq:approx2} .
\end{gather}
\end{subequations}
Hence, in this approximation, equations~(\ref{eq:approx1}) and (\ref{eq:approx2}) show that
\begin{equation}
 h^* =1 + O(\epsilon), \qquad  h/R = \epsilon + O(\epsilon)^2.
 \label{eq:film}
\end{equation}
The range of validity of such a narrow approximation can be estimated from the balance of the axial forces on the vesicle, requiring in some average sense, $\mathrm{d}p/\mathrm{d}x < 0$ and $\tau <0$. This means that the film thickness must lie in the following range:
\begin{subequations}\label{eq:val}
\begin{gather}
 2q/V < h < 3q/V, \label{eq:val1}\\
\mbox{or equivalently}  \qquad 1 - \frac{U}{V} < \delta \equiv h /R < \frac{3}{2} \left ( 1- \frac{U}{V} \right ) , \label{eq:val2} 
\end{gather} 
\end{subequations}
for the clearance parameter $\delta$. We will see that such conditions are always satisfied when $\lambda \to \lambda_c$.

Since we are mostly interested in the asymptotic behavior of quantities of interest established via a narrow-gap analysis, we consider the configuration close to maximal confinement (i.e., $\lambda \to \lambda_c$) in which a cylindrical vesicle with hemispherical ends is formed, nearly fitting the tube cross-section, as shown in Fig.~\ref{fig_app}. In this limit, a pure geometric consideration -- constraints of vesicle surface area and enclosed volume -- which are fixed, leads to an expansion for the clearance parameter $\delta$~\cite{barakat_shaqfeh_2018a}:
\begin{equation}
 \delta = 1 - \lambda/\lambda_c +  O\left [(1 - \lambda/\lambda_c)^2 \right].
 \label{eq:film2}
\end{equation}
Using Eq.~(\ref{eq:epsilon}) in Eq.~(\ref{eq:film}) gives an asymptotic behavior of the vesicle mobility, measured in the relative velocity 
\begin{equation}
 V/U = 1 + (1 - \lambda/\lambda_c) +  O\left [(1 - \lambda/\lambda_c)^2 \right].
 \label{eq:vel2}
\end{equation}
This prediction is the same as in the small-gap theory~\cite{barakat_shaqfeh_2018a} when~$\lambda \to \lambda_c$.

While these two asymptotic expansions are helpful to interpret the present numerical results regarding the film thickness and vesicle mobility, it remains unclear how the clearance parameter is precisely controlled dynamically by a quantity, such as hydrodynamic pressure $p$ in the lubrication layer or the membrane tension $\gamma$. To this end, by using (\ref{eq:approx}), we further simplify the normal and shear stress boundary conditions, equations~(\ref{lub:norm}) and (\ref{eq:tau}), which can be approximated by
\begin{subequations}\label{eq:simpli}
\begin{gather}
 p = \frac{\kappa}{2 R^3} - \frac{\gamma}{R} , \label{eq:simpli1}\\
 \frac{\mathrm{d} \gamma }{\mathrm{d}x} = -\tau =  -\frac{2\eta V}{h}  + \frac{6\eta q}{h^2} , \label{eq:simpli2} \\
 \mbox{with}  \qquad q = \left ( \frac{6\eta V}{h^2} +  \frac{\mathrm{d} p }{\mathrm{d}x} \right) \frac{h^3}{12 \eta}. \label{eq:simpli3}
\end{gather} 
\end{subequations}
We then obtain a simple expression for the membrane tension gradient in the lubrication layer region
\begin{equation}
\label{eq:film3}
\frac{\mathrm{d} \gamma }{\mathrm{d}x} = \frac{\eta V}{h} + O(1) ,
\end{equation}
thereby indicating that the thickness of the lubrication layer is inversely propositional to the tension gradient in the membrane. That equation gives 
\begin{equation}
\label{ }
\gamma(x) = \gamma_R + \frac{\eta V}{h} x
\end{equation}
with $\gamma_R$ denoting the membrane tension of the rear endcap. Therefore, the tension of the vesicle increases  linearly with distance and has a higher tension $ \gamma_F$ at the frontal endcap. The pressure in the lubrication layer, however, decreases with distance according to Eq.~(\ref{eq:simpli1}). The pressure and membrane tension both are of the order of $\epsilon^{-1}$ and, therefore bending resistance has a negligible contribution to the hydrodynamic force balance in the lubrication layer.  The rear tension $\gamma_R$ of the vesicle is negligibly small compared to its frontal counterpart $\gamma_F$~\cite{SecombSkakak1986,Bruinsma1996} -- the rear portion of the vesicle is nearly tensionless, we may estimate the frontal tension for the cylindrical portion having a length of $L_0$ ($ \simeq L - 2R$, when $\lambda \to \lambda_c$)
\begin{equation}
\label{eq:tension}
\gamma_F \simeq \frac{\eta V}{\delta} (L_0/R) .
\end{equation}
A further overall asymptotic solution of Eqs.~(\ref{eq:simpli}) near the front end of the vesicle has shown~\cite{SecombSkakak1986,Bruinsma1996} as
\begin{equation}
\label{eq:tension2}
\gamma_F \simeq \eta V \left (\delta/c_0\right)^{-3/2},
\end{equation}
where $c_0$ is a constant. Finally, the clearance parameter is found to be controlled through a dynamical parameter -- the vesicle tension-mobility-based capillary number $\mathrm{Ca_v}= \eta V/\gamma_F$
\begin{equation}
\label{eq:film4}
\delta  \simeq c_0 \mathrm{Ca^{2/3}_v}.
\end{equation}
The numerical prefactor $c_0$ differs slightly in the literature; $c_0 \simeq 2.123$ in Ref.~\cite{SecombSkakak1986}  while $c_0 \simeq 2.05$ in Ref.~\cite{Bruinsma1996}.

%\bibliography{biblio_soft_matter.bib}
%apsrev4-2.bst 2019-01-14 (MD) hand-edited version of apsrev4-1.bst
%Control: key (0)
%Control: author (8) initials jnrlst
%Control: editor formatted (1) identically to author
%Control: production of article title (0) allowed
%Control: page (0) single
%Control: year (1) truncated
%Control: production of eprint (0) enabled
%

\end{document}